\documentclass[12pt]{article}

\usepackage[margin=1in]{geometry} 
\usepackage{graphicx} 
\usepackage{amsmath} 
\usepackage{amsfonts} 
\usepackage{amsthm} 
\usepackage{xspace}
\usepackage{subfiles}
\usepackage{blindtext}
\usepackage{mymacros}
\usepackage{amssymb}
\usepackage{bm}
\usepackage{etoolbox}
\usepackage{marginnote}
\usepackage{tikz-cd}
\usepackage{pst-node}
\usepackage{todonotes}

\newtheorem{thm}{Theorem}[section]
\newtheorem{theorem}[thm]{Theorem}

\newtheorem{remark}[thm]{Remark}
\newtheorem{definition}[thm]{Definition}
\newtheorem{lem}[thm]{Lemma}
\newtheorem{lemma}[thm]{Lemma}
\newtheorem{prop}[thm]{Proposition}
\newtheorem{cor}[thm]{Corollary}

\begin{document}

\title{Symmetric Circuits for Rank Logic
   \thanks{Research funded in part by EPSRC grant EP/S03238X/1.  An extended abstract of this paper appeared in~\cite{DawarW18}}
}

\author{Anuj Dawar and Gregory Wilsenach \\
  University of Cambridge Computer Laboratory \\
  \texttt{firstname.lastname@cl.cam.ac.uk}}

\maketitle

\begin{abstract}
  Fixed-point logic with rank (FPR) is an extension of fixed-point logic with
  counting (FPC) with operators for computing the rank of a matrix over a finite
  field. The expressive power of FPR properly extends that of FPC and is
  contained in $\PT$, but it is not known if that containment is proper. We give
  a circuit characterization for FPR in terms of families of symmetric circuits
  with rank gates, along the lines of that for FPC given by [Anderson and Dawar
  2017]. This requires the development of a broad framework of circuits in which
  the individual gates compute functions that are not symmetric (i.e., invariant
  under all permutations of their inputs). This framework also necessitates the
  development of novel techniques to prove the equivalence of circuits and
  logic. Both the framework and the techniques are of greater generality than
  the main result.
\end{abstract}

\section{Introduction}\label{sec:introduction}

The study of extensions of fixed-point logics plays an important role in the
field of descriptive complexity theory. In particular, fixed-point logic with
counting ($\FPC$) has become a reference logic in the search for a logic for
polynomial-time (see~\cite{Dawar-siglog}). In this context, Anderson and
Dawar~\cite{AndersonD17} provide an interesting characterization of the
expressive power of $\FPC$ in terms of circuit complexity. They show that the
properties expressible in this logic are exactly those that can be decided by
polynomially-uniform families of circuits (with threshold gates) satisfying a
natural \emph{symmetry} condition. Not only does this illustrate the robustness
of $\FPC$ as a complexity class within $\PT$ by giving a distinct and natural
characterization of it, it also demonstrates that the techniques for proving
inexpressibility in the field of finite model theory can be understood as
lower-bound methods against a natural circuit complexity class. An interesting
example of the latter is the lower bounds for arithmetic circuits computing the
permanent obtained by these methods~\cite{DawarW20}. This raises an obvious
question (explicitly posed in the concluding section of~\cite{AndersonD17}) of
how to obtain circuit characterizations of logics more expressive than $\FPC$,
such as choiceless polynomial time (CPT) and fixed-point logic with rank
($\FPR$). It is this last question that we address in this paper.

Fixed-point logic with rank extends the expressive power of $\FPC$ by means of
operators that allow us to define the rank of a matrix over a finite field. Such
operators are natural extensions of counting---counting the dimension of a
definable vector space rather than just the size of a definable set. At the same
time they make the logic rich enough to express many of the known examples that
separate $\FPC$ from $\PT$. Rank logics were first introduced
in~\cite{Dawar09logicswith}. The version of $\FPR$ we consider here is that
defined by Gr\"adel and Pakusa~\cite{GradelP15a} where the prime characteristic
is a parameter to the rank operator, and we do not have a distinct operator for
each prime number. Formal definitions of these logics are given in
Section~\ref{sec:background}. We give a circuit characterization, in terms of
symmetric circuits, of $\FPR$. One might think, at first sight, that this is a
simple matter of extending the circuit model with gates for computing the rank
of a matrix. It turns out, however, that the matter is not so simple as the
symmetry requirement interacts in surprising ways with such rank gates. It
requires a new framework for defining classes of such circuits, which yields
remarkable new insights.

The word \emph{symmetry} is used in more than one sense in the context of
circuits (and also in this paper). We say that a Boolean function $f:\{0,1\}^n
\ra \{0,1\}$ is symmetric if the value of the function on a string $s$ is
determined by the number of $1$s in $s$. In other words, $f$ is invariant under
\emph{all} permutations of its input. In contrast, when we consider the input to
a Boolean function to be the adjacency matrix of an $n$-vertex graph, for
example, and $f : \{0,1\}^{n \choose 2} \ra \{0,1\}$ decides a graph property,
then $f$ is invariant under all permutations of its input induced by
permutations of the $n$ vertices of the graph. We call such a function
\emph{graph-invariant}. More generally, for a relational vocabulary $\tau$ and a
standard encoding of $n$-element $\tau$-structures as strings over $\{0,1\}$, we
can say that a function taking such strings as input is $\tau$-invariant if it
is invariant under permutations induced by the $n$ elements. A circuit $C$
computing such an invariant function is said to be \emph{symmetric} if every
permutation of the $n$ elements extends to an automorphism of $C$. It is
families of symmetric circuits in this sense that characterize $\FPC$
in~\cite{AndersonD17}. The restriction to symmetric circuits arises naturally in
the study of logics and has appeared previously under the names of generic
circuits in the work of~\cite{DENENBERG1986216} and explicitly order-invariant
circuits in the work of Otto~\cite{Otto1997}. In this paper, we use the word
``symmetric'', and context is used to distinguish the meaning of the word as
applied to circuits from its meaning as applied to Boolean functions.

The main result of~\cite{AndersonD17} says that the properties of
$\tau$-structures definable in $\FPC$ are exactly those that can be decided by
$\PT$-uniform families of symmetric circuits using AND, OR, NOT and majority
gates. Note that each of these gates itself computes a Boolean function that is
symmetric in the strong sense identified above. On the other hand, a rank
threshold function, e.g.\ one that takes as input an $n \times n$ matrix and
outputs $1$ if the rank of the matrix is greater than a threshold $t$, is
\emph{not} symmetric. In our circuit characterization of $\FPR$ we necessarily
have to consider such non-symmetric gates. Indeed, we can show that
$\PT$-uniform families of symmetric circuits with gates labelled by \emph{any}
symmetric functions do not take us beyond the power of $\FPC$. This is a further
illustration of the robustness of $\FPC$. In order to go beyond it, we need to
introduce gates for Boolean functions that are not symmetric. We construct a
systematic framework for including isomorphism-invariant functions that take
$\tau$-structures as inputs, for $\tau$ an arbitrary multi-sorted relational
vocabulary, in Section~\ref{sec:symm-circ}. We also explore what it means for
such circuits to be symmetric.

The proof of the circuit characterization of $\FPC$ relies on the \emph{support
  theorem} proved in~\cite{AndersonD17}. This establishes that for any
$\PT$-uniform family of circuits using $\AND$, $\OR$, $\NOT$ and majority gates
there is a constant $k$ such that every gate has a support of size at most $k$.
That is to say that we can associate with every gate $g$ in the circuit $C_n$
(the circuit in the family that works on $n$-element structures) a subset $X$ of
$[n]$ of size at most $k$ such that any permutation of $[n]$ fixing $X$
pointwise extends to an automorphism of $C_n$ that fixes $g$. This theorem is
crucial to the translation of the family of circuits into a formula of $\FPC$,
which is the difficult (and novel) direction of the equivalence. In attempting
to do the same with circuits that now use rank-threshold gates we are faced with
the difficulty that the proof of the support theorem in~\cite{AndersonD17}
relies in an essential way on the fact that the Boolean function computed at
each gate is symmetric. We are able to overcome this difficulty and prove a
support theorem for circuits with rank gates but this requires substantial,
novel technical machinery. This result is not only more general in that it's
applicability to circuits with rank gates, but is stronger even if we restrict
our attention to circuits with gates that only compute symmetric functions.

Another crucial ingredient in the proof of Anderson and Dawar is that we can
eliminate redundancy in the circuit $C_n$ by making it \emph{rigid}. That is, we
can ensure that the \emph{only} automorphisms of $C_n$ are those that are
induced by permutations of $[n]$. Here we face the difficulty that identifying
the symmetries and eliminating redundancy in a circuit that involves gates
computing $\tau$-invariant functions requires us to solve the isomorphism
problem for $\tau$-structures. This is a hard problem (or, at least, one that we
do not know how to solve efficiently) even when the $\tau$-structures are
$0$-$1$-matrices. We overcome this difficulty by placing a further restriction
on circuits that we call \emph{transparency}. Circuits satisfying this condition
have the property that their lack of redundancy is transparent.

In the characterization of $\FPC$, the translation from formulas into families
of circuits is easy and, indeed, standard. In our case, we have to show that
formulas of $\FPR$ translate into uniform families of circuits using
rank-threshold gates that are both symmetric and transparent. This is somewhat
more involved technically and presented in
Section~\ref{sec:formulas-to-circuits}. Finally, with all these tools in place,
the translation of such $\PT$-uniform families of circuits into formulas of
$\FPR$ given in Section~\ref{sec:circuits-to-formulas} completes the
characterization. This still requires substantial new techniques. The
translation of circuits to formulas in~\cite{AndersonD17} relies on the fact
that in order to evaluate a gate computing a symmetric Boolean function, it
suffices to count the number of inputs that evaluate to true and there is a
bijection between the orbits of a gate and tuple assignments to its support.
When counting is no longer sufficient, this bijection has to preserve more
structure and demonstrating this in the case of matrices requires new insight.

\section{Background}\label{sec:background}

We write $\nats$ to denote the positive integers and $\natz$ to denote the
non-negative integers. For $n \in \nats$ we write $[n]$ to denote the set $\{1,
\ldots, n\}$. For sets $X$ and $Y$ we write $X^{\underline{Y}}$ to denote the
set of injections from $X$ to $Y$. For $S \subseteq Y$ we let $f^{-1}[S] := \{x
\in X : f(x) \in S\}$ denote the inverse image of $S$. We say a relation $R
\subseteq \Pi_{j \in J} X_j$ is \emph{full} if $R = \Pi_{j \in J} X_j$.

\subsection{Group Theory}
For a set $S$ we write $\sym_S$ or $\sym(S)$ to denote the symmetric group on
$S$. For $n \in \nats$ we write $\sym_n$ to abbreviate $\sym_{[n]}$. Let $G$ be
a group acting on a set $X$ and let $S \subseteq X$. Let $\stab_G(S) := \{\pi
\in G : \forall s \in S , \,\, \pi(s) = s\}$ be the \emph{pointwise stabiliser}
of $S$. Let $\orb_G(x) := \{ y \in X : \exists \pi \in G , \,\, \pi(x) = y\}$ be
the \emph{orbit} of $x$. In both cases we omit subscripts when the group $G$ is
obvious from context. For $n \in \nats$ we write $\stab_n(S)$ to abbreviate
$\stab_{\sym_n}(S)$ and $\orb_n(x)$ to abbreviate $\orb_{\sym_n}(x)$.

Let $G$ be a group acting on a set $X$. We denote this as a left action, i.e.\
$\sigma x$ for $\sigma \in G$, $x \in X$. The action extends in a natural way to
powers of $X$. So, for $(x,y) \in X \times X$, $\sigma(x,y) = (\sigma x,\sigma
y)$. It also extends to the powerset of $X$ and functions on $X$ as follows. The
action of $G$ on $\pow(X)$ is defined for $\sigma \in G$ and $S \in \pow(X)$ by
$\sigma S = \{\sigma x : x \in S\}$. For any set $Y$, the action of $G$ on $Y^X$
is defined for $\sigma \in G$ and $f\in Y^X$ by $(\sigma f) (x) = f(\sigma x)$
for all $x \in X$. We refer to all of these as the \emph{natural action} of $G$
on the relevant set.

\subsection{Linear Algebra}
Let $A$, $B$, and $X$ be non-empty sets such that $A$ and $B$ finite. An $X$-valued
$A\times B$ \emph{matrix} is a function $M : A \times B \ra X$. If $X \subseteq
\natz$ we say $M$ is number valued.

We use $\ff$ to denote a field. For any prime $p$ we write $\ff_p$ for the
finite field of order $p$. For an $\ff_p$-valued matrix $M$ we write
$\rank_p(M)$ to denote the rank. For a number valued matrix $M$ and a prime $p$
we write $({M \bmod p})$ to denote the matrix formed by taking the residue of
each element modulo $p$. We write $\rank_p(M)$ to abbreviate $\rank_p ({M \bmod
  p})$.

\subsection{Logic}
A \emph{vocabulary} is a finite set of relation symbols $(R_1, \ldots, R_k)$,
each of which has a fixed \emph{arity}. For each relation symbol $R$ we write
$r_R$ to denote its arity. We omit the subscript when it is obvious from
context.

A \emph{many-sorted vocabulary} is a tuple of the form $(\bm{R}, \bm{S},
\zeta)$, where $\bm{R}$ is a relational vocabulary, $\bm{S}$ is a finite
sequence of \emph{sort} symbols, and $\zeta$ is a function that assigns to each
$R \in \bm{R}$ a tuple $\zeta(R) := (s_1, \ldots, s_{r})$ of sort symbols. We
call $\zeta(R)$ the \emph{type} of $R$. A \emph{$\tau$-structure} $\mathcal{A}$
is a tuple $(A , (R^{\mathcal{A}})_{R \in \bm{R}})$ where $A = \uplus_{s \in S }
A_{s}$ a disjoint union of non-empty sets, called the \emph{universe} of
$\mathcal{A}$, and for all $R \in \bm{R}$, $R^{\mathcal{A}} \subseteq U_{s_1}
\times \ldots \times U_{s_{r}}$, where $(s_1 , \ldots , s_{r_i}) = \zeta (R_i)$.
The \emph{size} of $\mathcal{A}$, denoted by $\vert \mathcal{A} \vert$, is the
cardinality of $A$. All structures in this paper are finite. Let $\fin{\tau}$
denote the set of all finite $\tau$-structures and for $n \in \nats$ let
$\fin{\tau, n}$ denote the set of all $\tau$-structures of size $n$. We say that
a $\tau$-structure is \emph{complete} if every relation it contains is full.

Let $\FO[\tau]$ denote \emph{first-order logic} over the vocabulary $\tau$. A
formula in $\FO[\tau]$ is formed from atomic formulas, each formed using
variables from some countable sequence of (first-order) variable symbols $x, y,
\ldots$, the relation symbols in $\tau$, and the equality symbol $=$, and then
closing the set of atomic formulas under the Boolean connectives, and universal
and existential quantification (i.e.\ $\land$, $\lor$, $\neg$, $\forall$, and
$\exists$). Let $\FP$ denote \emph{fixed-point logic}, the extension of $\FO$
with inflationary fixed-point operators. We assume standard syntax and semantics
for $\FO$ and $\FP$, and direct the reader to~\cite{grohe2017descriptive}
or~\cite{immerman1999descriptive} for more detail.

Let $L[\tau]$ be a logic over a vocabulary $\tau$ and $\phi \in L[\tau]$. For a
sequence of variables $\vec{x}$ we write $\phi(\vec{x})$ to denote that the free
variables in $\phi$ are among $\vec{x}$. For $\mathcal{A} \in \fin{\tau}$ and
$\vec{a} \in A^{\vert \vec{x} \vert}$ we write $\mathcal{A} \models_L \phi
[\vec{a}]$ to denote that $\phi$ holds when interpreted in $\mathcal{A}$ with
respect to the logic $L$ when each variable $x_i$ is assigned to $a_i$. We omit
the subscript $L$ when it is obvious from context.





\subsubsection{Fixed-Point with Counting}
Let $\FPC[\tau]$ denote \emph{fixed-point logic with counting} over the
vocabulary $\tau.$ $\FPC$ extends $\FP$ with a \emph{counting operator} that
allows us to define the cardinality of a definable set, as well as a number sort
for doing calculations. Each variable in $\FPC$ is either a \emph{number} or
\emph{element} variable. Element variables correspond to the variables in $\FO$
or $\FP$ and range over the universe of the structure. Number variables instead
range over the number domain $\natz$.

The number terms and formulas of $\FPC[\tau]$ are defined by mutual recursion. A
\emph{number term} is either a number variable, of the form $t_1 + t_2$ or $t_1
\cdot t_2$ for number terms $t_1, t_2$, or of the form $\# x \phi$, where $\phi
\in \FPC[\tau]$ and $x$ is an element variable. The number term $\# x \phi$
denotes the number of distinct elements $a$ for which $\phi[a]$ holds. Let
$t(\vec{z})$ be a number term. Let $\mathcal{A} \in \fin{\tau}$ and $\vec{a}$ be
a $\vert \vec{z} \vert$-sequence such that if $z_i$ is a element variable then
$a_i \in A$ and if $z_i$ is a number variable then $a_i \in \natz$. Let
$t^{\mathcal{A}}[\vec{a}]$ denote the number defined by evaluating $t$ in
$\mathcal{A}$ under the assignment that maps each $z_i$ to $a_i$.

The atomic formulas of $\FPC[\tau]$ include all atomic formulas in $\FP[\tau]$
as well as formulas of the form $t_1 = t_2$ and $t_1 \leq t_2$ for number terms
$t_1$ and $t_2$. The formulas of $\FPC[\tau]$ are formed by closing the set of
atomic formulas under the usual Boolean connectives, the first-order
quantifiers, and the fixed-point operator. To avoid undecidability,
quantification over the number sort must be bounded. If $\nu$ is a number
variable, it must be quantified in the form $\exists \nu \leq t \; \phi$ or
$\forall \nu \leq t \; \phi$ for $\phi \in \FPC[\tau]$, where $t$ is a number
term. We omit these bounds to abbreviate that the number variable is quantified
over $\{0, \ldots, \vert \mathcal{A} \vert\}$ when evaluated over some
$\mathcal{A} \in \fin{\tau}$. In other words, we write $\forall \nu \phi$ to
abbreviate $\forall \nu \leq t_m \phi$, where $t_m$ is a number term that
denotes the size of the structure over which it is evaluated. The fixed-point
operator may bind sequences of variables consisting of both element and number
variables. We similarly require an explicit bound for each number variable
quantified.

We assume a standard semantics of $\FPC[\tau]$ and direct the reader
to~\cite{grohe2017descriptive} for more detail.

Let $\natFP[\tau]$ denote \emph{fixed-point logic with a number sort} over the
vocabulary $\tau$. The formulas of $\natFP[\tau]$ include all those in
$\FPC[\tau]$ that do not include an application of the counting operator.

\subsubsection{Fixed-Point with Rank}
Let $\FPR[\tau]$ denote \emph{fixed-point logic with rank} over the vocabulary
$\tau$. $\FPR$ extends $\natFP$ with an operator that denotes the rank of a
definable matrix over a finite field. To define $\FPR$ we extend the definition
of $\natFP$ to include number terms of the form $[\rank (\vec{x} \vec{\mu} \leq
\vec{t}, \vec{y}\vec{\nu} \leq \vec{s}, \pi). \eta]$, where $\eta$ is a number
term and $\vec{t}$ and $\vec{s}$ are tuples of number terms bounding the
sequences of number variables $\vec{\mu}$ and $\vec{\nu}$, respectively.

Informally, this number term evaluates to the rank of the matrix $M$ over a
field with order determined by $\pi$, where $M$ is a number valued matrix with
entries determined by $\eta$ and with rows indexed by the assignments to
$\vec{x} \vec{\mu}$ and columns indexed by the assignments to
$\vec{y}\vec{\nu}$. More formally, we define the evaluation of this number term
for $\mathcal{A} \in \fin{\tau}$ as follows. Let $\natz^{\leq \vec{t}} :=
\{\vec{a} \in \natz^{\vert \vec{t} \vert} : \forall i \in [\vert \vec{t} \vert]
\,\,\ a_i \leq t^{\mathcal{A}}_i\}$ and let $\natz^{\leq \vec{s}}$ be defined
similarly. Let $R : = A^{\vert \vec{x} \vert} \times \natz^{\leq \vec{t} }$ and
$C := A^{\vert \vec{y} \vert} \times \natz^{\leq \vec{s}}$. Let $M_\eta : R
\times C \ra \natz$ be the matrix defined such that $M (\vec{a}\vec{m},
\vec{b}\vec{n}) = \eta^{\mathcal{A}}[\vec{a}\vec{m}, \vec{b}\vec{n}]$ for all
for $(\vec{a}\vec{m}, \vec{b}\vec{n}) \in R \times C$. Let $p =
\pi^{\mathcal{A}}$. The number term $[\rank (\vec{x} \vec{\mu} \leq \vec{t},
\vec{y}\vec{\nu} \leq \vec{s}, \pi). \eta]$ evaluates to $0$ if $p$ is not prime
and otherwise evaluates to $\rank_p(M_\eta)$.

For any formula $\phi$ the number term $[\rank(x, y, 2) (x = y \land \phi(x))]$
denotes the number of $a \in A$ such that $\mathcal{A} \models \phi[a]$. It
follows that $\FPR$ is at least as expressive as $\FPC$.

We should note that there are two different rank logics considered in the
literature. The first, introduced in~\cite{Dawar09logicswith}, is defined in a
similar manner as above but includes a separate rank operator for each prime,
rather then a single operator that takes in a prime as a parameter. This logic
was shown to be strictly contained in $\PT$~\cite{GradelP15a}. The rank logic we
have defined here is strictly more expressive and is the focus of current
research. For more details on the syntax and semantics of $\FPR$
see~\cite{GradelP15a}.

\section{Symmetric Circuits}\label{sec:symm-circ}

We aim to derive a characterisation for $\FPR$ analogous to the characterisation of
$\FPC$ in terms of symmetric majority circuits given in~\cite{AndersonD17}. A
natural starting point would be to consider circuits defined over a basis more
expressive than the majority basis. However, in
Section~\ref{sec:limitations-symmetric-basis} we show that any basis of
symmetric functions leads to a circuit model that is at most as expressive as
$\FPC$. Since $\FPR$ is strictly more expressive than $\FPC$ we need to consider
circuits with gates labelled by non-symmetric functions.

This poses a potential problem as the usual notion of a circuit as a directed
acyclic graph implicitly requires that the gates be labelled only by symmetric
functions. This follows from the fact that a directed acyclic graph imposes no
structure on the inputs of any gate, and so each gate must compute a function
invariant under any ordering of its inputs, i.e.\ a symmetric function.
Importantly, the standard basis, majority basis, and most other bases considered
in the literature satisfy this assumption.

In this section we introduce the notion of a \emph{structured function}, a
Boolean function whose input strings naturally encode relational structures. We
generalise the notion of a symmetric function for this framework, and so
introduce \emph{isomorphism-invariant structured functions}. We then define a
more general circuit model so as to allow for gates to be labelled by
isomorphism-invariant structured functions. We generalise the notions of a
circuit automorphism and a symmetric circuit for this new model. We also address
many problems that emerge in the analysis of these more general circuits, and
discuss a connection with the graph isomorphism problem.

\subsection{Structured Functions}
\label{sec:structured-functions}

We now define structured functions formally.

\begin{definition}
  Let $\tau := (\bm{R}, \bm{S}, \zeta)$ be a vocabulary and let $X = \uplus_{s
    \in \bm{S}} X_s$ be a disjoint union of finite non-empty sets. Let $K$ be
  the complete $\tau$-structure with universe $X$. Let $\tot{\tau}{X} :=
  \uplus_{R \in \bm{R}}R^{K}$. We call a function $F : \{0,1\}^{\tot{\tau}{X}}
  \ra \{0,1\}$ a \emph{structured function}.
\end{definition}

Let $F: \{0,1\}^{\tot{\tau}{X}} \ra \{0,1\}$ be a structured function. We call
$\tau$ the \emph{vocabulary} of $F$, $X$ the \emph{universe} of $F$, $K$ the
\emph{structure} associated with $F$, and $\tot{\tau}{X}$ the \emph{index} of
$F$. We denote the index of $F$ by $\ind(F)$, the universe of $F$ by
$\universe{F}$, and the structure associated with $F$ by $\str{F}$. We abuse
notation and write $(x, R) \in \tot{\tau}{X}$ to denote that $x \in
\tot{\tau}{X}$ and $x \in R^{K}$. In other words, $(x, R) \in \tot{\tau}{X}$
denotes that $x$ is a tuple from the relation $R$ in the complete
$\tau$-structure with universe $X$. Throughout this paper we assume that each sort symbol
appears in the type of some relation, i.e.\ for each sort $s \in \tau$ there
exists a relation symbol $R \in \tau$ such that $s$ appears in $\zeta(R)$.

We identify elements in $\{0, 1\}^{\tot{\tau}{X}}$ with $\tau$-structures over
$X$. More precisely, we identify each $f \in \{0,1\}^{\tot{\tau}{X}}$ with the
$\tau$-structure $\mathcal{A}_f$ with universe $X$ such that for each relation
symbol $R \in \tau$, $R^{\mathcal{A}_f} = \{x: (x, R) \in \tot{\tau}{X}, f(x) =
1\}$.

The action of $\sym(\tot{\tau}{X})$ on $\tot{\tau}{X}$ extends naturally to
$\{0,1\}^{\tot{\tau}{X}}$ and further to structured functions of the form $\{0,
1\}^{\tot{\tau}{X}} \ra \{0, 1\}$. For a function $F: \{0, 1\}^{\tot{\tau}{X}}
\ra \{0, 1\}$ the stabiliser group $\stab(F)$ is the set of all permutations
$\sigma \in \sym(\tot{\tau}{X})$ such that for all $x \in \tot{\tau}{X}$, $F
(\sigma x) = F (x)$. Let $H := \bigoplus_{s \in \bm{S}}\sym_{X_s}$ and consider
$H$ as a subgroup of $\sym(\tot{\tau}{X})$. Then $H \leq \stab(F)$ if, and only
if, $F$ is invariant under isomorphism. This motivates the following definition.

\begin{definition}
  Let $\tau := (\bm{R}, \bm{S}, \zeta)$ be a vocabulary and let $X = \uplus_{s
    \in \bm{S}} X_s$ be a disjoint union of finite non-empty sets. Let $F : \{0,
  1\}^{\tot{\tau}{X}} \ra \{0,1\}$ be a structured function. We call $H :=
  \bigoplus_{s \in \bm{S}}\sym_{X_s}$ the \emph{natural invariance group} of
  $F$. We say that $F$ is \emph{isomorphism invariant} if $H \leq \stab(F)$.
\end{definition}

We write $\natinv(F)$ to denote the natural invariance group of $F$. We identify
the usual symmetric Boolean functions with isomorphism-invariant structured
functions defined over a vocabulary consisting of a single unary relation. We
call these functions \emph{fully symmetric} to distinguish them from
isomorphism-invariant structured functions, which are symmetric in a weaker
sense.

We now introduce some notation and a few useful conventions. Throughout this
paper we can often assume, without loss of generality, that a structured
function has as its universe a direct sum of initial segments of the natural
numbers. We introduce some notation to simplify the definition of such a
function. Let $\tau= (\bm{R}, \bm{S}, \zeta)$ be a many-sorted relational
vocabulary. Let $e : \bm{S} \ra \nats$. We write $\tot{\tau}{e}$ to denote
$\tot{\tau}{X}$ where $X = \uplus_{s \in \bm{S}}X_s$, for each $s \in \bm{S}$,
$X_s = [e(s)]$. If $\bm{S} = \{s_1, \ldots, s_m\}$ then for $p_1, \ldots, p_m
\in \nats$ we write $\tot{\tau}{p_1, \ldots, p_m}$ to denote $\tot{\tau}{e}$
where $e : \bm{S} \ra \nats$ is defined by $e(s_i) = p_i$ for each $i \in [m]$.

We now generalise the notion of a basis to this framework.

\begin{definition}
  A \emph{Boolean basis} (or just a \emph{basis}) is a set of
  isomorphism-invariant structured functions.
\end{definition}

We now define the \emph{rank threshold functions}. Let $\matvoc$ be a three
sorted vocabulary with a single three-sorted ternary relation symbol $R$. We
first define a translation between number-valued matrices and
$\matvoc$-structures. Let $S$ be a $\matvoc$-structure with universe $X =
\uplus_{s \in [3]} X_s$. Let $M_S : X_1 \times X_2 \ra \nats$ be defined for
$(x_1, x_2) \in X_1 \times X_2$ by
\begin{align*}
  M_S (x_1, x_2) = \vert \{x_3 \in X_3 : R^S(x_1, x_2, x_3)\}\vert.
\end{align*}
We call $S$ a \emph{three sorted matrix} and $M_S$ the \emph{matrix associated
  with} $S$. For a prime $p$ we write $\rank_p(S)$ to abbreviate $\rank_p(M_S)$.

Let $t, p \in \natz$ and $X = \uplus_{s \in [3]}X_s$ be a disjoint union of
non-empty sets. Let $\RANK^t_p[X] : \{0,1\}^{\tot{\matvoc}{X}} \ra \{0,1\}$ be
defined for each $S \in \{0,1\}^{\tot{\matvoc}{X}}$ by
\begin{align*}
  \RANK^t_p[X] (S) = \begin{cases} 1 &\quad \text{if } p  \text{ prime and } \rank_p(S) \geq t \\  0 &\quad \text{otherwise}\\ \end{cases}.
\end{align*}
These functions are isomorphism-invariant. We write
$\RANK^t_p[a, b, c]$ for $a, b, c \in \nats$ to denote $\RANK^t_p[X]$ where $X$
is the disjoint union of $[a]$, $[b]$, and $[c]$. We omit $X$ and just write
$\RANK^t_p$ when $X$ is clear from context.

For $n \in \nats$ we similarly define $\AND[n] : \{0,1\}^n \ra \{0,1\}$ to be
the usual $\AND$ function on $n$ bits, $\OR[n]$ to be the $\OR$ function on $n$
bits, $\NOT$ to be negation on 1 bit, and $\MAJ[n]$ the majority function on $n$
bits. All of these functions are fully symmetric.

The \emph{standard basis}, denoted by $\BS$, contains $\NOT$ and all $\AND[n]$
and $\OR[n]$ for $n \in \nats$. The \emph{majority basis}, denoted by $\BM$,
extends the standard basis with the majority functions, i.e.\ $\MAJ[n]$ for all
$n \in \nats$. The \emph{rank basis}, denoted by $\BR$, is the extension of
$\BS$ with the rank threshold functions, i.e.\ $\RANK^t_p[a, b, c]$ for all $ a,
b, c, t, p \in \nats$

\subsection{Symmetric Circuits}
\label{sec:symmetric-circutis}
We now generalise the circuit model of Anderson and Dawar~\cite{AndersonD17} so
as to allow for circuits to be defined over bases of isomorphism-invariant
functions, rather than just fully symmetric functions. In this model each gate
$g$ is not only associated with an element of the basis, as in the conventional
case, but also with a labelling function. This labelling function maps the input
gates of $g$ to an appropriate set of labels (i.e.\ the index of the structured
function associated with $g$). In concord with this generalisation, we also
update the circuit-related notions discussed by Anderson and
Dawar~\cite{AndersonD17}, e.g.\ circuit automorphisms, symmetry, etc. Moreover,
we briefly discuss some of the important complications introduced by our
generalisation, and introduce some of the important tools we use later to
address these complications.

\begin{definition}[Circuits on Structures]
  Let $\mathbb{B}$ be a basis and $\rho$ be a relational vocabulary, we define a
  \emph{$(\mathbb{B}, \rho)$-circuit} $C$ of order $n$ computing a $q$-ary query
  $Q$ as a structure $\langle G, \Omega, \Sigma, \Lambda, L \rangle$.
  \begin{myitemize}
    \setlength\itemsep{0mm}
  \item $G$ is called the set of gates of $C$.
  \item $\Omega$ is an injective function from $[n]^q$ to $G$. The gates in the
    image of $\Omega$ are called the output gates. When $q = 0$, $\Omega$ is a
    constant pointing to a single output gate.
  \item $\Sigma$ is a function from $G$ to $\mathbb{B} \uplus \rho \uplus
    \{0,1\} $ such that $\vert \Sigma^{-1} (0) \vert \leq 1$ and $\vert
    \Sigma^{-1} (1) \vert \leq 1$. Those gates mapped to $\rho \uplus \{0,1\}$
    are called input gates, with those mapped to $\rho$ called relational gates
    and those mapped to $\{0,1\}$ called constant gates. Those gates mapped to
    $\mathbb{B}$ are called internal gates.
  \item $\Lambda$ is a sequence of injective functions $(\Lambda_{R})_{R \in
      \rho}$ such that $\Lambda_{R}$ maps each relational gate $g$ with $\Sigma
    (g) = R$ to the tuple $\Lambda_{R} (g) \in [n]^{{\arityr{R}}}$. When no
    ambiguity arises we write $\Lambda (g)$ for $\Lambda_{R} (g)$.
  \item $L$ associates with each internal gate $g$ a function $L(g):
    \ind(\Sigma(g)) \rightarrow G$ such that if we define a relation $W
    \subseteq G^{2}$ by $W(h_1,h_2)$ iff $h_2$ is an internal gate and $h_1$ is
    in the image of $L(h_2)$, then $(G, W)$ is a directed acyclic graph.
  \end{myitemize}
\end{definition}

The definition requires some explanation. Each gate in $G$ computes a function
of its inputs and the relation $W$ on $G$ is the set of ``wires''. That is,
$W(h,g)$ indicates that the value computed at $h$ is an input to $g$. However,
since the functions are structured, we need more information on the set of
inputs to $g$ and this is provided by the labelling $L$. For any $g$,
$\Sigma(g)$ tells us what the function computed at $g$ is, and thus the index of
$\Sigma(g)$ tells us the structure on the inputs and $L(g)$ maps this to the set
of gates that form the inputs to $g$.

Let $C := \langle G, \Omega, \Sigma, \Lambda, L \rangle$ be a $(\mathbb{B},
\rho)$-circuit of order $n$. We define the \emph{size} of $C$, denoted $\vert C
\vert$ to be the number of elements in $G$. If $(C_n)_{n \in \nats}$ is a family
of circuits we assume that each $C_n$ is a circuit of order $n$.

For each $g \in G$ we let $W(\cdot, g) := \{h \in G : W(h,g)\}$ and $W(g, \cdot)
:= \{h \in G : W(g,h)\}$. We call the elements of $W(\cdot, g)$ the
\emph{children} of $g$ and the elements of $W(g, \cdot)$ the \emph{parents} of
$g$. We also abbreviate $W(g, \cdot)$ by $H_g$. We write $W_T$ for the
transitive closure of $W$. We say $g$ is an \emph{input gate} if it is not an
internal gate. The \emph{depth} of $g$ is $0$ if $g$ is an input gate or
otherwise the length of the longest path from an input gate to $g$.

For a gate $g \in G$ with $\Sigma(g) \in \mathbb{B}$, we let the \emph{index} of
$g$, denoted by $\ind(g)$, be the index set of $\Sigma(g)$. We abuse notation
and let the \emph{vocabulary} and \emph{universe} of $g$ denote the vocabulary
and universe of $\Sigma(g)$, respectively. We also use $\vocab{g}$ and
$\universe{g}$ to denote the vocabulary and universe of $g$, respectively, and
write $\natinv(g)$ to abbreviate $\natinv(\Sigma(g))$. We write $\str{g}$ to
denote the structure associated with $\Sigma(g)$. We say that a gate $g$ is
\emph{fully symmetric} if $\Sigma(g)$ is fully symmetric. We say $C$ is a
\emph{circuit with fully symmetric gates} if every gate in $C$ is fully
symmetric.
  
Let $\rho$ be a relational vocabulary, $\mathcal{A}$ be a $\rho$-structure with
universe $A$ of size $n$, and $\gamma \in [n]^{\underline{A}}$. Let $\gamma
\mathcal{A}$ be the structure with universe $[n]$ formed by mapping the elements
of $A$ in accordance with $\gamma$. The evaluation of a $(\mathbb{B},
\rho)$-circuit $C$ of order $n$ computing a $q$-ary query $Q$ proceeds by
recursively evaluating the gates in the circuit. The evaluation of the gate $g$
for the bijection $\gamma$ and input structure $\mathcal{A}$ is denoted by
$C[\gamma \mathcal{A}](g)$, and is given as follows:
\begin{myenum}
\item If $g$ is a constant gate then it evaluates to the value given by
  $\Sigma(g)$;
\item if $g$ is a relational gate then $g$ evaluates to $1$ iff $\gamma
  \mathcal{A} \models \Sigma(g)(\Lambda (g))$; and
\item if $g$ is an internal gate let $L^{\gamma \mathcal{A}}(g): \ind(g)
  \rightarrow \{0,1\}$ be defined by $L^{\gamma\mathcal{A}}(g)(x) = C[\gamma
  \mathcal{A}](L(g)(x))$, for all $x \in \ind(g)$. Then $g$ evaluates to $1$ iff
  $\Sigma(g) (L^{\gamma \mathcal{A}}(g)) = 1$.
\end{myenum}
We say that $C$ defines the $q$-ary query $Q \subseteq A^q$ under $\gamma$ where
$\vec{a} \in Q$ if, and only if, $C[\gamma \mathcal{A}](\Omega (\gamma \vec{a}))
= 1$. We write $C[\gamma \mathcal{A}]$ to denote the query $Q$. In general the
evaluation $C$ depends on the particular encoding of $\mathcal{A}$ as a
structure with universe $[n]$, i.e.\ on the chosen bijection $\gamma$. We
consider circuits whose outputs do not depend on this choice of encoding.
Anderson and Dawar~\cite{AndersonD17} call such a circuit \emph{invariant}. We
have reproduced their definition below.

\begin{definition}[Invariant Circuit]
  Let $C$ be a $(\mathbb{B}, \rho)$-circuit of order $n$, computing some $q$-ary
  query. We say $C$ is \emph{invariant} if for every $\rho$-structure
  $\mathcal{A}$ of size $n$, $\vec{a} \in A^{q}$, and $\gamma_1, \gamma_2 \in
  [n]^{\underline{A}}$ we have that $C[\gamma_1 \mathcal{A}](\Omega (\gamma_1
  \vec{a})) = C[\gamma_2 \mathcal{A}](\Omega (\gamma_2 \vec{a}))$.
\end{definition}

If a family of $(\mathcal{B}, \rho)$-circuits $\mathcal{C}$ is invariant it
follows that the query computed is a $q$-ary query on $\rho$-structures. When $q
= 0$ then $\mathcal{C}$ computes a property of $\rho$-structures, i.e.\ a
decision problem. Each circuit $C_n \in \mathcal{C}$ then defines a structured
function with vocabulary $\rho$ and universe $[n]$.

\begin{lem}
  Let $C$ be a $(\mathbb{B}, \rho)$-circuit of order $n$ computing a $0$-ary
  query. The structured function computed by $C$ is isomorphism-invariant if,
  and only if, $C$ is an invariant circuit.
\end{lem}
\begin{proof}
  Let $F_C$ be the structured function defined by $C$ that maps
  $\rho$-structures over $[n]$ to $\{0, 1\}$. Suppose $F_C$ is
  isomorphism-invariant. Let $\mathcal{A}$ be a $\rho$-structure of size $n$ and
  let $\gamma_1, \gamma_2 \in [n]^{\underline{A}}$. Then $C[\gamma_1
  \mathcal{A}](\Omega) = F_C(\gamma_1\mathcal{A}) = F_C(\gamma_2\mathcal{A}) =
  C[\gamma_2 \mathcal{A}](\Omega)$. The second equality follows from the fact
  that $\gamma_1 \mathcal{A}$ and $\gamma_2\mathcal{A}$ are isomorphic. So $C$
  is invariant.

  Suppose $C$ is invariant. Let $\mathcal{B}$ be a $\rho$-structure over $[n]$.
  Then $[n]^{\underline{B}}$ is the set of permutations of $[n]$. Let $\sigma, e
  \in \sym_n$, where $e$ is the identity. Then $F_C(\mathcal{B}) = C[e
  \mathcal{B}](\Omega) = C[\sigma \mathcal{B}](\Omega) = F_C(\sigma \mathcal{B})
  = (\sigma F_C)(\mathcal{B})$. So $F_C$ is isomorphism-invariant.
\end{proof}

We now define an automorphism of a circuit, generalising the definition
introduced by Anderson and Dawar. The definition is similar, but adds the
requirement that if a gate $g$ is mapped to $g'$, then children of $g$ must be
mapped to the children of $g'$ via some appropriate isomorphism of the structure
associated with $g$.

\begin{definition}[Automorphism]\label{defn:automorphism}
  Let $C = \langle G, \Omega, \Sigma, \Lambda, L\rangle$ be a
  $(\mathbb{B},\tau)$-circuit of order $n$ computing a $q$-ary query. Let
  $\sigma \in \sym_n$ and $\pi: G \rightarrow G$ be a bijection such that
  \begin{myitemize}
  \item for all output tuples $x \in [n]^q$, $\pi \Omega (x) = \Omega (\sigma
    x)$,
  \item for all gates $g \in G$, $\Sigma (g) = \Sigma (\pi g)$,
  \item for each relational gate $g \in G$, $\sigma \Lambda (g) = \Lambda (\pi
    g)$, and
  \item For each pair of gates $g, h \in G$ we have $W(h,g)$ if, and only if,
    $W(\pi h, \pi g)$ and for each internal gate $g$ we have that $L(\pi g)$ and
    $ \pi L(g)$ are isomorphic (as labelled structures).
  \end{myitemize}
  We call $\pi$ an \emph{automorphism} of $C$, and we say that $\sigma$
  \emph{extends to an automorphism} $\pi$. The group of automorphisms of $C$ is
  denoted by $\aut (C)$.
\end{definition}
We can equally define an isomorphism between a pair of circuits $C = \langle G,
\Omega, \Sigma, \Lambda, L\rangle$ and $C' = \langle G', \Omega', \Sigma',
\Lambda', L'\rangle$ as a bijection $\pi: G \rightarrow G'$ satisfying
conditions as above. We do not usually need to consider distinct, isomorphic
circuits and for this reason we only formally define automorphisms.

We are particularly interested in circuits that have the property that
\emph{every} permutation in $\sym_n$ extends to an automorphism of the circuit.

\begin{definition}[Symmetry]
  A circuit $C$ or order $n$ is called \emph{symmetric} if every $\sigma \in
  \sym_n$ extends to an automorphism on $C$.
\end{definition}

It follows that for any symmetric circuit $C$ of order $n$ there is a
homomorphism $h$ that maps $\sym_n$ to $\aut(C)$ such that if $\sigma \in
\sym_n$ then $h(\sigma)$ is an automorphism extending $\sigma$. If $C$ does not
contain a relational gate then it computes a constant function. These circuits
form a trivial and uninteresting class and so, unless stated otherwise, we
assume each circuit in this paper contains at least one relational gate. With
this assumption in place it follows that some element of $[n]$ appears in a
tuple labelling some relational gate, and so by symmetry every element of $[n]$
appears in a tuple labelling a relational gate. Then no two distinct elements of
$\sym_n$ agree on all input gates, and so the homomorphism $h$ is injective.

Suppose $h$ is not also surjective. Since every automorphism extends some
permutation, there exists $\sigma \in \sym_n$ and $\pi, \pi' \in \aut(C)$ both
of which extend $\sigma$ but disagree on some gate $g$. It follows from
Lemmas~\ref{lem:syntactic-equivalence-equal-function}
and~\ref{lem:permutation-extending-syntactic-equivalence} that, although
$\pi(g)$ and $\pi'(g)$ are not equal, they are in a precise sense ``essentially
the same''. In this way the existence of a non-surjective $h$ witnesses the
introduction of ``artefacts'' into our analysis. We therefore restrict ourselves
to circuits with \emph{unique extensions}, a property which ensures that $h$ is
an isomorphism.

\begin{definition}
  We say that a circuit $C$ over order $n$ has \emph{unique extensions} if for
  every $\sigma \in \sym_n$ there is at most one $\pi \in \aut(C)$ such that
  $\pi$ extends $\sigma$.
\end{definition}

Most of the important technical tools needed in this paper are only applicable
to circuits with unique extensions. A similar problem arises in the work of
Anderson and Dawar~\cite{AndersonD17}. They address this by defining a normal
form for circuits, which they call \emph{rigid circuits}, and show that these
circuits have unique extensions. They then show that symmetric circuits can be
transformed into rigid symmetric circuits in polynomial time, allowing them to
restrict their attention to circuits with unique extensions (and numerous other
desirable properties) without loss of generality.

We will define a normal form analogous to rigidity and show that there is
polynomial time algorithm that transforms a circuit into an equivalent circuit
of this form. It is at this point that we arrive at the first complication
introduced by our generalisation. The polynomial-time translation
in~\cite{AndersonD17} makes indispensable use of the polynomial-time
decidability of many important circuit properties for circuits defined over
bases of fully symmetric functions. However, for the more general circuits
discussed here, it is not known if even the most basic circuit properties are
polynomial-time decidable. This is essentially due to the requirement built in
to Definition~\ref{defn:automorphism} that an automorphism that takes $g$ to
$g'$ must be an isomorphism between $L(g)$ and $L(g')$. This makes checking the
condition as hard as isomorphism checking. As such, constructing an argument
analogous to~\cite{AndersonD17}, as well as establishing the numerous other
crucial results whose proofs rely on the polynomial-time decidability of various
circuit properties, would be beyond the scope of this paper.

In order to proceed we explicitly restrict our attention to a particular class
of circuits characterised by a restriction on the children of gates labelled by
functions that are not fully symmetric. We say such circuits are
\emph{transparent}. We show in Section~\ref{sec:algorithms} that all of the
circuit properties of interest are polynomial-time decidable for transparent
circuits, and we use these results to define a polynomial-time transformation
from transparent circuits to equivalent circuits in our normal form. This normal
form not only allows us to ensure the polynomial-time decidability of numerous
circuit properties but also ensures the circuits have unique extensions, a
necessary requirement to apply the theory of supports we develop in
Section~\ref{sec:symm-support}. Importantly, while the restriction to
transparent circuits makes it easier to translate families of circuits into
formulas, the usual translation from formulas to circuits does not produce a
family of transparent circuits in general. We discuss this difficulty and define
a novel translation from formulas to symmetric circuits that ensures
transparency in Section~\ref{sec:formulas-to-circuits}.

Before we can formally define \emph{transparency} we need to define the
\emph{syntactic equivalence} relation on the gates of a circuit. The intuition
is that two gates $g$ and $g'$ in a circuit are syntactically equivalent if the
circuits underneath them are ``hereditarily equivalent'', i.e.\ if the two
circuits induced by the restrictions to $W_T (\cdot, g)$ and $W_T(\cdot, g')$
are in some sense just copies of one another.

\begin{definition}
  Let $C := \langle G, \Omega, \Sigma, \Lambda, L \rangle$ be a $(\mathbb{B},
  \rho)$-circuit of order $n$. Let $\equiv'$ be an equivalence relation on on
  $G$ defined such that for $g, h \in G$, $g \equiv' h$ if, and only if, $\Sigma
  (g) = \Sigma (h)$ and either both $g$ and $h$ are not output gates or $g$ and
  $h$ are both output gates and $g = h$. We define now define a refinement
  $\equiv$ of $\equiv'$ by induction on depth. Let $g, h \in G$ be such that $g
  \equiv' h$. If $g$ and $h$ have depth $0$ then $g \equiv h$ if, and only if,
  both $g$ and $h$ are constant gates or both $g$ and $h$ are relational gates
  and $\Lambda (g) = \Lambda (h)$. Let $r > 0$ and suppose we have defined
  $\equiv$ for the case that both $g$ and $h$ have depth less than $r$. If
  either $h$ or $g$ have depth $r$ then $g \equiv h$ if, and only if, there
  exists $\lambda \in \natinv(g)$ such that for all $x \in \ind(g)$,
  $L(g)(\lambda (x)) \equiv L(h)(x)$. We call the relation $\equiv$
  \emph{syntactic equivalence}.
\end{definition}

For a circuit $C$ of order $n$ and a gate $g$ we think of $g$ as computing the
function that maps an input structure $\mathcal{A}$ and a bijection $\gamma$
from the universe of $\mathcal{A}$ to $[n]$ to the evaluation $C[\gamma
\mathcal{A}](g)$. While we would like to be able to identify gates that compute
the same function in this sense, it is not hard to show that deciding this
equivalence relation for a given circuit is $\NP$-hard. We now show that if two
gates are syntactically equivalent then the functions computed at these two
gates must be equal. We show later that the syntactic equivalence relation is
polynomial-time decidable for the class of circuits of interest to us. In this
sense we shall treat syntactic equivalence as a tractable refinement of an
$\NP$-complete relation.

\begin{lem}
  Let $C = \langle G, \Omega, \Sigma, \Lambda, L \rangle$ be a $(\BB,
  \rho)$-circuit of order $n$. Let $\mathcal{A}$ be a $\rho$-structure of size
  $n$ and let $\gamma$ be a bijection from the universe of $\mathcal{A}$ to
  $[n]$. For all $g, g' \in G$ if $g \equiv g'$ then $C[\gamma \mathcal{A}](g) =
  C[\gamma \mathcal{A}](g')$.
  \label{lem:syntactic-equivalence-equal-function}
\end{lem}
\begin{proof}
  We prove the result by induction on depth. Suppose $g$ and $g'$ have depth $0$
  and $g \equiv g'$. Then they are both input gates and so $g = g'$. The result
  for depth $0$ then follows trivially. Suppose $g$ and $g'$ are internal gates,
  and suppose for all $h, h' \in G$ of depth less than $g$ or $g'$ we have that
  if $h \equiv h'$ then $C[\gamma \mathcal{A}](h) = C[\gamma \mathcal{A}](h')$.
  Suppose $g \equiv g'$. There exists $\lambda \in \natinv(g)$ such that $L(g)
  (x) \equiv L(g') (\lambda x)$ for all $x \in \ind(g)$. It follows from the
  inductive hypothesis that $L^{\gamma \mathcal{A}}(g)(x) = C[\gamma
  \mathcal{A}](L(g)(x)) = C[\gamma \mathcal{A}](L(g')(\lambda x)) = (L^{\gamma
    \mathcal{A}}(g) \lambda) (x)$ for all $x \in \ind(g)$. Since $\Sigma(g)$
  (and so $\Sigma(g')$) is a structured function, it follows that $C[\gamma
  \mathcal{A}](g) = \Sigma(g)(L^{\gamma \mathcal{A}} (g)) = \Sigma(g')
  (L^{\gamma \mathcal{A}}(g') \lambda) = \Sigma(g') (L^{\gamma \mathcal{A}}(g'))
  = C[\gamma \mathcal{A}](g')$. The result follows.
\end{proof}

The syntactic equivalence relation identifies gates that have ``equivalent''
circuits underneath them. A similar intuition is captured by identifying gates
that are mapped to one another by automorphisms that extend the trivial
permutation. We now show that if two automorphisms extend the same permutation
then the two images of each gate must be syntactically equivalent.

\begin{lem}
  Let $C$ be a circuit of order $n$, $\sigma \in \sym_n$ and $\pi, \pi' \in
  \aut(C)$ both extend $\sigma$, then for every gate $g$ in the circuit we have
  that $\pi (g)$ and $\pi'(g)$ are syntactically equivalent.
  \label{lem:permutation-extending-syntactic-equivalence}
\end{lem}
\begin{proof}
  From the definition of an automorphism we have for any gate $g$ in $C$ that
  $\Sigma (g) = \Sigma (\pi (g)) = \Sigma (\pi' (g))$, and either all of $g$,
  $\pi (g)$ and $\pi'(g)$ are not output gates or all three are output gates and
  $\Omega (\Omega^{-1}(g)) = \Omega (\sigma \Omega^{-1}(g)) = \pi' \Omega
  (\Omega^{-1}(g)) = \pi'(g)$.
  
  We now prove the result by induction on depth. Suppose $g$ is a gate of depth
  $0$. Then $g$ is either a relational or constant gate. In either case $\pi (g)
  = \pi' (g) = g$, and so $\pi(g)$ and $\pi'(g)$ are syntactically equivalent.

  Suppose $g$ is an internal gate and suppose that for every gate $h$ of depth
  less than $g$, $\pi (h)$ is syntactically equivalent to $\pi'(h)$. We have
  that there exists $\lambda, \lambda' \in \natinv(g)$ such that $\pi L(g)(x) =
  L(\pi g) (\lambda x)$ and $\pi' L(g)(x) = L(\pi' g)(\lambda' x)$, for all $x
  \in \ind(g)$. Then, from the inductive hypothesis, we have that $\pi L(g)(x)
  \equiv \pi 'L(g)(x)$ and so $L(\pi g)(\lambda x) \equiv L(\pi' g) (\lambda'
  x)$, for all $x \in \ind(g)$. Thus we have that $L(\pi g) (x) = L (\pi g)
  (\lambda \lambda^{-1}(x)) \equiv L(\pi' g) (\lambda' \lambda^{-1}(x))$, for
  all $x \in \ind(g)$, and so $L(\pi g)/_\equiv$ is isomorphic to $L(\pi'g)
  /_\equiv$. We thus have that $\pi (g) $ and $\pi' (g)$ are syntactically
  equivalent, and the result follows.
\end{proof}

It follows from Lemma~\ref{lem:permutation-extending-syntactic-equivalence} that
the syntactic equivalence relation of a circuit $C$ constrains the automorphism
group of $C$ and the orbits and stabiliser groups of the gates in $C$. It
follows then that if all of the internal gates of $C$ have trivial syntactic
equivalence classes, i.e.\ for all $g$ and $h$ in $C$ if $g \equiv h$ then $g =
h$, then $C$ has unique extensions. This motivates the importance of the
following definition.

\begin{definition}
  We say that a circuit $C$ is \emph{reduced} if $C$ has trivial syntactic
  equivalence classes.
\end{definition}

We now define the notion of a transparent circuit, as well as numerous other
restrictions on the structure of a circuit.

\begin{definition}
  Let $C$ be a circuit and $g$ be a gate in $C$. We say $g$ has \emph{injective
    labels} if $L(g)$ is an injection. We say $g$ has \emph{unique children} if
  no two distinct gates in $H_g$ are syntactically equivalent. We say $g$ has
  \emph{unique labels} if $g$ has injective labels and unique children.

  We say $C$ has \emph{injective labels} (or just $C$ is \emph{injective}) if
  every gate in $C$ has injective labels. We say $C$ has \emph{unique labels} if
  every gate in $C$ has unique labels. We say $C$ is \emph{transparent} if every
  gate $g$ in $C$ that is not fully symmetric has unique labels.
\end{definition}

We earlier noted the need for a normal form analogous to the notion of a rigid
circuit introduced by Anderson and Dawar~\cite{AndersonD17}. We will show that
the notion of a reduced injective circuit suffices as our normal form. We have
already noted that these circuits have unique extensions. This allows us to
apply the theory of supports developed in Section~\ref{sec:symm-support}.
Moreover, we show in Section~\ref{sec:algorithms} that there is a polynomial
time translation from transparent circuits to reduced injective circuits. We
also show that many natural circuit properties can be computed in polynomial
time for such circuits. These results together play a crucial role in our
translation from families of circuits to formulas in
Section~\ref{sec:circuits-to-formulas}.

We say a circuit $C$ is a \emph{rank circuit} if it is defined over the basis
$\BR$. We are now in a position to state the main theorem of this paper.

\begin{thm}[Main Theorem]
  Let $\rho$ be a relational vocabulary and let $Q$ be a $\rho$-query. Then $Q$
  is definable in $\FPR[\rho]$ if, and only if, $Q$ is definable by a
  $\PT$-uniform family of transparent symmetric rank circuits with vocabulary
  $\rho$.
  \label{thm:main-result}
\end{thm}

\subsection{Limitations of Symmetric Bases}
\label{sec:limitations-symmetric-basis}
The main result of this subsection, Theorem~\ref{thm:symmetric-circuits-bound},
establishes that any symmetric circuit defined over an arbitrary basis of fully
symmetric functions can be transformed in polynomial time into a
symmetric majority circuit computing the same function. It follows from this
result that, so long as we are interested in circuit families of at least
polynomial size, it makes no difference to the expressive power of the model if
we consider symmetric majority circuits or symmetric circuits defined over the
basis containing \emph{all} fully symmetric functions. Moreover, in order to
define a symmetric circuit characterisation for $\FPR$ analogous to the
symmetric circuit characterisation of $\FPC$ we must consider bases that include
functions that are not fully symmetric.

We first show that any fully symmetric function can be computed by a small
symmetric majority circuit. The crucial observation used is that for any
fully symmetric function $F : \{0,1\}^n\rightarrow \{0,1\}$ and input
$\vec{x}$ the value of $F(\vec{x})$ is entirely determined by the number of $1$s
in $\vec{x}$. In particular, $F$ is characterised by the set $c_{F} \subseteq
[n]$ such that $m \in c_F$ if, and only if, there exists $\vec{x} \in \{0,
1\}^n$ with $m$ $1$s such that $F(\vec{x}) = 1$. We encode $F$ by the set $c_F$.

\begin{prop}
  \label{prop:fuctions-maj}
  There is a linear time algorithm that takes as input a fully symmetric 
  function $F: \{0,1\}^n \rightarrow \{0,1\}$ and outputs a symmetric majority
  circuit $C$ that computes $F$ and has depth at most $5$, width at most $2n+2$
  and size at most $5n + 3$.
\end{prop}

\begin{proof}
  We have $c_{F}$ from the input. We now define $C$. We define the set of gates
  and wires of $C$ layer by layer as follows. The first layer consists of the
  input gates $x_1, \ldots, x_n$. The second layer consists of two $\MAJ$ gates
  for each $a \in c_{F}$, which we denote by $\maj_a$ and $\maj^{\neg}_a$. For
  each $a \in c_{F}$ there is one wire from each of $x_1, \ldots , x_n$ to
  $\maj_a$ and $\maj^{\neg}_a$. For all $a \geq \frac{n}{2}$, there are $2a - n$
  wires from $0$ to $\maj_a$ and $2a - n + 2$ wires from $0$ to $\maj^{\neg}_a$.
  For all $a < \frac{n}{2}$ there are $n - 2a$ wires from $1$ to $\maj_a$ and $n
  - 2a - 2$ wires from $1$ to $\maj^{\neg}_a$. The third layer consists of one
  $\NOT$ gate for each $a \in c_{F}$, which we denote by $\neg_a$. For each $a
  \in c_{F}$ there is a wire from $\maj^\neg_a$ to $\neg_a$. The fourth layer
  consists of one $\AND$ gate for each $a \in c_{F}$, which we denote by
  $\countgate_a$. For each $a \in c_{F}$ there is a wire from each of $\maj_a$
  and $\neg_a$ to $\countgate_a$. The fifth layer consists of just a single
  $\OR$ gate, designated as the output gate, and for each $a \in c_{F}$ there is
  a wire from $\countgate_a$ to the output gate.

  We summarise the circuit up to the fourth layer as follows:
  \[
    \countgate_a = \begin{cases} \land ( \maj ( x_1, \ldots, x_n, \underbrace
      {0, \ldots, 0}_{2a - n} ), \neg ( \maj (x_1, \ldots , x_n, \underbrace{0,
        \ldots,
        0}_{2a - n + 2} ) )) &  a \geq \frac{n}{2} \\
      \land ( \maj ( x_1, \ldots, x_n, \underbrace {1, \ldots, 1}_{n - 2a} ),
      \neg ( \maj ( x_1, \ldots , x_n, \underbrace{1, \ldots, 1}_{n - 2a -2} )
      )) & a < \frac{n}{2}.
    \end{cases}
  \]
  For an input $\vec{x} \in \{0, 1\}^n$, $\countgate_a$ evaluates to $1$ if, and only
  if, the number of $1$s in $\vec{x}$ equals $a$. Thus $C$ evaluates to $1$ if,
  and only if, there exists $a \in c_{F}$ such that the number of $1$s in
  $\vec{x}$ equals $a$ if, and only if, $F (\vec{x}) = 1$.

  Let $\sigma \in \sym_n$. We define the automorphism $\pi$ extending $\sigma$
  as follows. If $x_i$ is an input gate let $\pi (x_i) := x_{\sigma (i)}$. For
  each majority gate $g$ in the second layer there is exactly one wire from each
  input gate to $g$. We note additionally that every other gate in the circuit
  is connected to the (non-constant) input gates only through a gate in the
  second layer. As such, if $g$ is an internal gate we let $\pi (g) := g$. Then
  $\pi$ is an automorphism extending $\sigma$, and so $C$ is symmetric. It is
  easy to see that the construction of the circuit $C$ can be implemented by an
  algorithm running in time polynomial in $n$.

  The first layer contains $n$ gates and the second layer contains at most $2
  \vert c_{F} \vert \leq 2n$ gates. The third and fourth layer each contain at
  most $n$ gates. The size of $C$ is at most $n + 2n + 2n + 1 + 2 = 5n +3$ (the
  additional 2 is for the constant gates), the width of $C$ is at most $2n + 2$,
  and the depth is at most $5$.
\end{proof}

\begin{thm}
  Let $\BB$ be a basis of fully symmetric functions. There is an algorithm
  that takes as input a symmetric circuit over $\BB$ and outputs a symmetric
  majority circuit computing the same function. This algorithm runs in time
  polynomial in the size of the input circuit.
  
  \label{thm:symmetric-circuits-bound}
\end{thm}

\begin{proof}
  Let $C$ be a symmetric circuit over $\BB$. We construct the corresponding
  symmetric majority circuit $C'$ by replacing each gate labelled by a fully symmetric function $F$ with the corresponding circuit given in
  Proposition~\ref{prop:fuctions-maj}. It is easy to see that $C'$ is symmetric
  and computes the same function as $C$.

\end{proof}


\section{Symmetry and Support}\label{sec:symm-support}

In this section we develop a theory of supports for symmetric circuits. We begin
with the definition of a support.

\begin{definition}
  Let $G \leq \sym_n$ and let $S \subseteq [n]$. Then $S$ is a \emph{support} of
  $G$ if $\stab_n(S) \leq G$.
\end{definition}

Let $g$ be a gate in a circuit $C$ of order $n$. We say $S \subseteq [n]$ is a
\emph{support} of $g$ if $S$ is a support of $\stab_n(g)$. In this section we
establish a bound on the minimum-size support of $g$. We also extend
the definition of a support and this bound to elements of the universe of a
gate. These results are used in Section~\ref{sec:circuits-to-formulas} for
defining the translation from $\PT$-uniform families of injective reduced
symmetric rank circuits to $\FPR$-formulas.

We now show that so long as $G \leq \sym_n$ has at least one support that is
small enough then it has a unique minimal support. We first state a useful lemma on
supports.

\begin{lem}[{\cite[Lemma 26]{BLASS1999141}}]
  Let $G \leq \sym_n$ and let $A, B \subseteq [n]$ be supports of $G$. If $A
  \cup B \neq [n]$ then $A \cap B$ is a support of $G$.
  \label{lem:inter-supports}
\end{lem}

\begin{lem}
  Let $G \leq \sym_n$ and suppose $G$ has a support of size less than
  $\frac{n}{2}$. Then there exists a unique minimum size support of $G$.
  \label{lem:minimum-support}
\end{lem}
\begin{proof}
  Let $S := \bigcap \{T \subseteq [n] : T \text{ supports } G \text{ and } \vert
  T\vert < \frac{n}{2}\}$. By repeated application of
  Lemma~\ref{lem:inter-supports} it follows that $S$ is a support of $G$. Let $T
  \subseteq [n]$ be a support of $G$ such that $\vert T \vert \leq \vert S
  \vert$. Then $\vert T \vert < \frac{n}{2}$ and so $S \subseteq T$ and thus $S
  = T$.
\end{proof}

Let $G \leq \sym_n$ have a support of size less than $\frac{n}{2}$. We call the
unique minimum-size support of $G$ the \emph{canonical support} of $G$ and denote it
by $\consp(G)$.

\subsection{The Support Theorem}
Let $g$ be a gate in a symmetric circuit $C$ with unique extensions of order
$n$. If $g$ has a support of size less than $\frac{n}{2}$ then it follows from
Lemma~\ref{lem:minimum-support} that $g$ has a unique minimum-size support. We
call this the support the \emph{canonical support} of $g$ and denote it by
$\consp(g)$.

In this subsection we state and prove the \emph{support theorem}. This
essentially says that when the orbits in a circuit $C$ have size bounded by a
polynomial, then every gate in $C$ has a support bounded in size by a constant.
To get to a proof of the theorem, we use a general result about permutation
groups. To understand this, suppose $G \leq \sym_n$ is a group with index
$[\sym_n:G] \leq n^k$. This is the case, for instance, for $G = \stab(g)$ when
the orbit of $g$ has size at most $n^k$. One way to have such a group $G$ of
small index is when $G$ contains a large symmetric group. That is $G$ contains
\emph{all} permutations on $n-k$ elements, or in other words $G$ has a support
of size $k$ or less. These are clearly not the only subgroups of $\sym_n$ of
small index as witnessed by the alternating group $\alt_n$.
Theorem~\ref{thm:dixonmort} below tells us that this is, essentially, the only
counter-example. That is, if $G$ has small index in $\sym_n$, then it must
contain a large alternating group. We then show that, in the case of circuits,
this counter-example does not arise and the stabilizer groups of gates always
contain a large symmetric group.

\begin{thm}[\cite{dixon1996permutation}, Theorem 5.2B]
  Let $Y$ be a set such that $n := \vert Y \vert > 8$, and let $k$ be an integer
  with $1 \leq k \leq \frac{n}{4}$. Suppose that $G \leq \sym_Y$ has index
  $[\sym_Y:G] < {n \choose k}$ then for some $X \subseteq Y$ with $\vert X \vert
  < k$ we have $\stab_{\alt_Y}(X) \leq G$.
  \label{thm:dixonmort}
\end{thm}

The second result needed to prove the support theorem asserts that if $g$ is
moved by some permutation $\sigma$ then there are two children of $g$ that
\emph{witness} this fact. In other words, there exist gates $h, h' \in H_g$ such
that $\sigma$ either moves $h$ or $h'$ outside of $H_g$ or $\sigma$ moves $h$
and $h'$ so as to ensure that the structures on the children of $g$ and $\sigma
g$ cannot be isomorphic. We now define this notion formally.

\begin{definition}
  Let $C$ be an injective symmetric circuit with unique extensions of order $n$,
  let $g$ be a gate in $C$. Let $h, h' \in H_g$ and let $\sigma \in \sym_n$. We
  say that $(h, h')$ \emph{witnesses that $\sigma$ moves $g$} if at least one of
  the following hold:
  \begin{itemize}
  \item $\sigma h \not\in H_g$,
  \item $\sigma h' \not\in H_g$, or
  \item for every $\lambda \in \natinv(g)$ either $\lambda L(g)^{-1}(h) \neq
    L(g)^{-1}(\sigma h)$ or $\lambda L(g)^{-1}(h') \neq L(g)^{-1}(\sigma h')$.
  \end{itemize}
\end{definition}

We now show that for any permutation $\sigma \in \sym_n$, $\sigma$ moves an
internal gate $g$ if, and only if, there exists a pair of children of $g$ that
witness that $\sigma$ moves $g$.

\begin{lem}
  Let $C$ be an injective symmetric circuit with unique extensions of order $n$,
  let $g$ be an internal gate in $C$, and let $\sigma \in \sym_n$. Then $\sigma
  \in \stab_n(g)$ if, and only if, for all $h, h' \in H_g$, $(h, h')$ does not
  witnesses that $\sigma$ moves $g$.
  \label{lem:witness}
\end{lem}
\begin{proof}
  $\Rightarrow$: Suppose $\sigma \in \stab_n(g)$. Then $\sigma H_g = H_g$ and
  there exists $\lambda$ such that $\sigma L(g)(x) = L(g)(\lambda x)$ for all $x
  \in \ind(g)$. It follows that for all $h, h' \in H_g$ we have that $\sigma
  L(g)(L(g)^{-1}(h)) = L(g)(\lambda L(g)^{-1}(h))$ and so $L(g)^{-1}(\sigma h) =
  \lambda L(g)^{-1}(h)$ and similarly $L(g)^{-1}(\sigma h') = \lambda
  L(g)^{-1}(h')$.

  $\Leftarrow$: Suppose for all $h, h' \in H_g$ we have that $(h, h')$ does not
  witnesses that $\sigma$ moves $g$. It follows that $\sigma H_g = H_g$. Let
  $\lambda : \universe{g} \ra \universe{g}$ be defined by $\lambda (a) =
  (L(g)^{-1}(\sigma (L(g)(\vec{a}))))(i)$ for all $a \in \universe{g}$ where
  $(\vec{a}, R) \in \ind(g)$ and $i \in [\vert \vec{a} \vert]$ are such that
  $\vec{a}(i) = a$. We now show that $\lambda$ is well defined. Let $(\vec{a}_1,
  R), (\vec{a}_2, R) \in \ind(g)$ and let $i$ and $j$ be such that $a :=
  \vec{a}_1(i) = \vec{a}_2(j)$. Since $(L(g)(\vec{a}_1), L(g)(\vec{a}_2))$ does
  not witness that $\sigma$ moves $g$, it follows that there exists $\lambda'
  \in \natinv(g)$ such that $\lambda' (\vec{a}_1) = \lambda'
  (L(g)^{-1}(L(g)(\vec{a}_1))) = L(g)^{-1}(\sigma (L(g)(\vec{a}_1)))$ and
  similarly $\lambda' (\vec{a}_2)) = L(g)^{-1}(\sigma (L(g)(\vec{a}_2)))$. Then
  \begin{align*}
    \lambda(a) &= L(g)^{-1}(\sigma (L(g)(\vec{a}_1)))(i)= (\lambda' (\vec{a}_1,)))(i) = (\lambda' (\vec{a}_2)))(j) \\&= L(g)^{-1}(\sigma (L(g)(\vec{a}_2)))(j) = \lambda(a).
  \end{align*}
  It follows almost immediately from the definition of $\lambda$ that $\lambda
  \in \natinv(g)$ and $L(g)\lambda = \sigma L(g)$, and so $\sigma \in
  \stab_n(g)$.
\end{proof}

\begin{cor}
  Let $C$ be an injective symmetric circuit with unique extensions of order $n$
  and let $g$ be an internal gate in $C$. Let $h, h' \in H_g$, $\sigma \in
  \sym_n$, and $\pi \in \stab_n(h) \cap \stab_n(h')$. If $(h, h')$ witnesses
  that $\sigma$ moves $g$ then $\sigma \pi \not\in \stab_n(g)$.
  \label{cor:preserve-witness}
\end{cor}
\begin{proof}
  Let $h, h' \in H_g$, $\sigma \in \sym_n$, and $\pi \in \stab_n(h) \cap
  \stab_n(h')$. Suppose $(h, h')$ witnesses that $\sigma$ moves $g$. From
  Lemma~\ref{lem:witness} it follows that either (i) $\sigma \pi h = \sigma h
  \not\in H_g$, (ii) $\sigma \pi h' = \sigma h' \not\in H_g$, or (iii) for every
  $\lambda \in \natinv(g)$ either $\lambda L(g)^{-1}(\pi h) = \lambda
  L(g)^{-1}(h) \neq L(g)^{-1}(\sigma h) = L(g)^{-1}(\sigma \pi h)$ or
  (similarly) $\lambda L(g)^{-1}(\pi h') \neq L(g)^{-1}(\sigma \pi h')$. Thus
  $(\pi h, \pi h')$ witnesses that $\sigma \pi$ moves $g$, and so from
  Lemma~\ref{lem:witness} $\sigma \pi \not\in \stab_n(g)$.
\end{proof}

Let $C$ be a symmetric circuit of order $n$. For each gate $g \in C$ let
$\SPs{g}$ be the minimal size of a support of $g$. Let $\SPs{C} := \max
\{\SPs{g} : g \in C\}$ and $\ORB{C} := \max_{g \in C}\vert \orb(g) \vert$.

We are now ready to prove the support theorem.

\begin{thm}[Support Theorem]
  Let $C$ be an injective symmetric circuit with unique extensions of order $n >
  8$. For every $1 \leq k \leq \frac{n}{4}$ if $\ORB{C} < {n \choose k}$
  then $\SPs{C} < k$.
  \label{thm:support-theorem}
\end{thm}

\begin{proof} Let $n > 8$ and let $k \in [\frac{n}{4}]$ be such that $\ORB{C} <
  {n \choose k}$. We prove by induction on the structure of the circuit that
  each gate has a support of size less than $k$.

  Suppose $g$ is an input gate in $C$. If $g$ is a constant gate then the empty
  set is a support. Suppose $g$ is a relational gate. Let $S_g$ be the set of
  distinct elements in $\Lambda(g)$. This is clearly a support of $g$ and also
  $\stab_n (g) = \stab_n(S_g)$. From the orbit-stabiliser theorem we have
  $\frac{n!}{(n - \vert S_g \vert)!} = [\sym_n : \stab_n(g)] = \ORB{g} < {n
    \choose k} = \frac{n!}{k! (n - k)!}$. It follows that $k! (n - k)! < (n -
  \vert S_g \vert)!$ and so $\vert S_g \vert < k$.
  
  Suppose $g$ is an internal gate and for all $h \in H_g$ we have $\SPs{h} < k$.
  From Theorem~\ref{thm:dixonmort} it follows that there exists $X \subseteq
  [n]$ such that $\vert X \vert < k$ and $\stab_{\alt_n}(X) \leq \stab_{n}(g)$.
  Suppose, for the sake of contradiction, that $\stab_n (X) \not\leq
  \stab_n(g)$. Then for all $a, b \in [n] \setminus X$ we have $\trans{a}{b}
  \not\in \stab_n(g)$, as if $\trans{a}{b} \in \stab_n(g)$ then $\stab_{n}(X)$
  is generated by $\stab_{\alt_n}(X) \cup \{\trans{a}{b}\}$ and is thus a
  subgroup of $\stab_n(g)$.

  Let $a, b \in [n] \setminus X$ be distinct. Then $\trans{a}{b} \not\in
  \stab_n(g)$. It follows from Lemma~\ref{lem:witness} that there exists $h_1,
  h_1' \in H_g$ such that $(h_1, h_1')$ witnesses that $\trans{a}{b}$ moves $g$.
  Since $3k + 2 < n$ there exist distinct $c, d \in [n] \setminus (X \cup
  \consp(h_1) \cup \consp(h_1'))$. Then $\trans{c}{d}$ moves $g$ and fixes both
  $h_1$ and $h_1'$. From Corollary~\ref{cor:preserve-witness} it follows
  $\trans{a}{b}\trans{c}{d} \not\in \stab_n(g)$. This yields the desired
  contradiction as $\trans{a}{b}\trans{c}{d} \in \stab_{\alt_n}(X) \leq
  \stab_n(g)$.
\end{proof}

\begin{remark}
  As noted in Section~\ref{sec:introduction}, Otto~\cite{Otto1997} previously
  introduced and studied explicitly order-invariant circuits, a model very
  similar to the symmetric circuits discussed here. The main lemma of his paper
  is a result analogous to Theorem~\ref{thm:support-theorem} (see~\cite[Lemma
  9]{Otto1997}). His proof is short and straight-forward, and has the
  considerable advantage of not relying on other results from the literature --
  as ours does. Otto's techniques can be adapted to prove a support theorem
  similar to Theorem~\ref{thm:support-theorem}. However, the resultant theorem
  places a lower upper bound on the value of $k$ and so, while it would suffice
  to prove our main theorem, it is a weaker result.
  \label{rem:support-theorem-otto}
\end{remark}

The following corollary establishes that if a family of injective symmetric
circuits with unique extensions has a polynomial size bound on the sizes of the
orbits then there is a constant bound on the sizes of the supports.

\begin{cor}
  Let $(C_n)_{n \in \nats}$ be a family of injective symmetric circuits with
  unique extensions and such that $\ORB{C_n} = \mathcal{O}(n^k)$ for some $k$.
  Then $\SPs{C_n} = \mathcal{O}(1)$.
  \label{cor:poly-size-support-bound}
\end{cor}
\begin{proof}
  Let $k \in \nats$ be such that $\ORB{C_n} = \mathcal{O}(n^k)$. Then $\ORB{C_n}
  = \mathcal{O}({n \choose k+1})$. It follows from
  Theorem~\ref{thm:support-theorem} that for large enough $n$, $\SPs{C_n}$ is
  bounded by some constant multiple of $k+1$, and so $\SPs{C_n} =
  \mathcal{O}(1)$.
\end{proof}



\subsection{Supports and Indexes}
We now extend our discussion of supports to include not only gates but elements
of the universes of gates. We first define for each gate $g$ in a symmetric
circuit a natural action of $\spstab{g}$ on the universe of $g$. This gives rise
to a similar notion of a support for these elements. We show that the support
theorem extends to this context.

Let $C$ be an injective symmetric circuit with unique extensions of order $n$,
let $g$ be an internal gate in $C$ with a support of size less than
$\frac{n}{2}$, and let $a \in \universe{g}$. We define an action of $\spstab{g}$
on the universe of $g$ by $\sigma \cdot a := (L(g)^{-1} \sigma L(g)
(\vec{a}))(i)$, for $\sigma \in \spstab{g}$, $a \in \universe{g}$, and where
$(\vec{a}, R) \in \ind(g)$ and $i \in [\vert \vec{a} \vert]$ is such that $a =
\vec{a}(i)$.

We say that $S \subseteq [n]$ is a \emph{support of $a$ in $g$} if $\stab(S)
\leq \stab_{\spstab{g}}(a)$. It follows from Lemma~\ref{lem:minimum-support}
that if there is a support of $a$ in $g$ of size less than $\frac{n}{2}$ then
there is a unique minimum-size support of $a$ in $g$. We call this minimum-size
support the \emph{canonical support of $a$ in $g$} and denote it by
$\consp_g(a)$. We write $\SPu{g}{a}$ to denote the minimum size of a support of
$a$ in $g$ and $\uSP{C}$ to denote the maximum of $\SPu{g}{a}$ for any gate $g$
and $a \in \universe{g}$.

We aim to extend the support theorem to elements of the universes of the gates
in a circuit. We first show that the canonical support of an element of the
universe of a gate $g$ is contained in the union of the canonical supports of
$g$ and a child of $g$.

\begin{lem}
  Let $C$ be an injective symmetric circuit with unique extensions of order $n$
  such that $\SPs{C} \leq \frac{n}{2}$. Let $g$ be a gate in $C$ and let $a \in
  \universe{g}$. Let $h \in H_g$ be such that there exists $(\vec{a}, R) \in
  \ind(g)$ where $a \in \vec{a}$ and $L(g)(\vec{a}) = h$. Then $\consp(h)\cup
  \consp(g)$ is a support of $a$ in $g$.
  \label{lem:bound-sup-universe}
\end{lem}
\begin{proof}
  Note that $\stab_{\spstab{g}}(h) = \bigcap_{b \in
    L(g)^{-1}(h)}\stab_{\spstab{g}}(b)$. Then
  \begin{align*}
    \stab(\consp(h) \cup \consp(g)) &\leq \stab(\consp(h)) \cap \stab(\consp(g)) \\&\leq \stab (h) \cap \spstab{g} \\ &=  \stab_{\spstab{g}}(h) \\ &=  \bigcap_{b \in
                                                                                                                                                     L(g)^{-1}(h)}\stab_{\spstab{g}}(b) \\ &\leq \stab_{\spstab{g}}(a).
  \end{align*}
\end{proof}

The following extension of the support theorem is an immediate corollary of
Lemma~\ref{lem:bound-sup-universe} and Theorem~\ref{thm:support-theorem}.

\begin{thm}
  Let $C$ be an injective symmetric circuit with unique extensions of order $n >
  8$. For every $1 \leq k \leq \frac{n}{4}$ if $\ORB{C} < {n \choose k}$ then
  $\uSP{C} < 2k$ and $\SPs{C} < k$.
  \label{thm:support-theorem-universe}
\end{thm}
\begin{proof}
  Let $g$ be a gate in $C$ and Let $h \in H_g$ be such that there exists
  $(\vec{a}, R) \in \ind(g)$ where $a \in \vec{a}$ and $L(g)(\vec{a}) = h$. It
  follows from Theorem~\ref{thm:support-theorem} that $\vert \consp(h) \vert <
  k$ and $ \vert \consp(g) \vert < k$. From Lemma~\ref{lem:bound-sup-universe}
  it follows that $\consp(h) \cup \consp(g)$ is a support of $a$ in $g$.
\end{proof}

We can derive from Theorem~\ref{thm:support-theorem-universe} a corollary
analogous to Corollary~\ref{cor:poly-size-support-bound}.

\begin{cor}
  Let $(C_n)_{n \in \nats}$ be a family of injective symmetric circuits with
  unique extensions such that $n \mapsto \ORB{C_n} = \mathcal{O}(n^k)$ for some
  $k$. Then $n \mapsto \uSP{C_n} = \mathcal{O}(1)$.
  \label{cor:poly-size-support-elements-bound}
\end{cor}

\section{Normal Forms and Decidability}\label{sec:algorithms}

In this section we show that many important circuit parameters can be computed
in polynomial time so long as we restrict our attention to transparent circuits.
These results are used in Section~\ref{sec:circuits-to-formulas} to establish a
translation from $\PT$-uniform families of transparent symmetric rank circuits
to $\FPR$-formulas.

The first result we prove, Lemma~\ref{lem:transparent-syntactic-equiv},
establishes that the syntactic equivalence relation for transparent circuits can
be computed in polynomial time. We use this result to show in
Proposition~\ref{prop:transparent-polynomial-time} that the transparency
property itself is polynomial-time decidable.

We show in Lemma~\ref{lem:transparent-unique} that there is a polynomial-time
algorithm that takes as input a transparent circuit $C$ and outputs a reduced
injective circuit computing the same function as $C$. This result allows us to
assume, without a loss of generality, that all $\PT$-uniform families of
transparent symmetric circuits consist of reduced injective circuits.

In the remainder of this section we consider the problem of computing supports
and orbits. We show in Lemma~\ref{lem:compute-automorphisms} that there is a
polynomial-time algorithm that determines the image of a gate in a reduced
circuit under the action of a given permutation. We use this algorithm to show
in Lemma~\ref{lem:computing-support-orbit} that there exists a polynomial-time
algorithm that determines if a reduced circuit is symmetric, and if so computes
the canonical supports and orbits of the gates in that circuit. We conclude this
section by showing that this result can be extended to the support and orbits of
elements of the universe of a gate.

We now show that the syntactic equivalence relation can be computed in
polynomial time for transparent circuits.

\begin{lem}
  There is an algorithm that takes as input a transparent circuit $C$ and
  outputs the syntactic equivalence relation on the gates of $C$. The algorithm
  runs in time polynomial in the size of $C$.
  \label{lem:transparent-syntactic-equiv}
\end{lem}
\begin{proof}
  The syntactic equivalence relation is defined inductively, and this definition
  can be implemented as an algorithm. To show that this algorithm runs in time
  polynomial in the size of the circuit it suffices to show that there exists a
  polynomial time algorithm that takes as input a transparent circuit $C$, two
  gates $g$ and $h$, and the syntactic equivalence relation for all gates of
  depth less than either $g$ or $h$ and decides if there exists $\lambda \in
  \natinv(h)$ such that for all $x \in \ind(h)$, $L(h)(\lambda(x)) \equiv
  L(g)(x)$.
  
  This algorithm is given as follows. We first attempt to greedily construct an
  bijection $f : \ind(g) \ra \ind(h)$ such that for all $x \in \ind(g)$,
  $L(g)(x) \equiv L(h)(f(x))$. If this construction fails halt and output that
  $g$ and $h$ are not syntactically equivalent. We check if $f$ is an
  isomorphism from the structure associated with $g$ to the structure associated
  with $h$. If so we output that $g \equiv h$ and otherwise output that $g
  \not\equiv h$.
\end{proof}

We now show that transparency is polynomial-time decidable. The algorithm works
by first checking that each gate that is not fully symmetric has injective
labels and then iterating over the circuit layer by layer, where a layer is a
set of gates of the same depth, at each step computing the syntactic equivalence
relation for the gates visited so far using
Lemma~\ref{lem:transparent-syntactic-equiv} and then using this to check whether
there exists a gate that is not fully symmetric and without unique labels in the
next layer. This algorithm can be easily implemented and we omit a full proof.

\begin{prop}
  There is an algorithm that takes as input a circuit and decides if that
  circuit is transparent. This algorithm runs in time polynomial in the size of
  the circuit.
  \label{prop:transparent-polynomial-time}
\end{prop}

We now define what it means to take a quotient of a circuit by the syntactic
equivalence relation. Intuitively, the quotient of a circuit is defined by
``merging'' each syntactic equivalence class into a single gate and including a
wire between any two gates $[h]$ and $[g]$ in the quotient circuit if, and only
if, there is a wire from $h$ to $g$.

\begin{definition}
  Let $C := \langle G, \Omega, \Sigma, \Lambda, L \rangle$ be a $(\BB,
  \rho)$-circuit. A \emph{quotient of $C$} is a $(\BB, \rho)$-circuit $C_\equiv
  := \langle G_\equiv , \Omega_\equiv, \Sigma_\equiv , \Lambda_\equiv, L_\equiv
  \rangle$, where $G_\equiv = G /_\equiv$, $\Omega_\equiv = \Omega /_\equiv$,
  $\Sigma = \Sigma /_\equiv$, $(\Lambda_\equiv)_R = \Lambda_R /_\equiv$ for all
  $R \in \rho$, and for all $[g] \in G_\equiv$ there exists $g' \in [g]$ such
  that $L_\equiv([g]) = L(g')/_\equiv$.
\end{definition}

Note that if $C$ and $C'$ are quotients of the same circuit then for each gate
$g$, $L(g)$ and $L'(g)$ are isomorphic as labelled structures. It follows that
$C$ and $C'$ are isomorphic in the precise sense alluded to right after
Definition~\ref{defn:automorphism}.

We now show that taking the quotient of a circuit preserves important
properties, including the function computed by the circuit, the symmetry of the
circuit, and whether the circuit has unique labels. We also show that the
quotient of a circuit is reduced.

\begin{lem}
  Let $C := \langle G, \Omega, \Sigma, \Lambda, L\rangle$ be a $(\BB,
  \rho)$-circuit and $C_{\equiv} = \langle G_\equiv , \Omega_\equiv,
  \Sigma_\equiv , \Lambda_\equiv, L_\equiv \rangle$ be a quotient of $C$. Then
  $C_\equiv$ is reduced and $C$ and $C_{\equiv}$ compute the same function.
  Moreover, if $C$ is symmetric then $C_{\equiv}$ is symmetric, and for all $g
  \in G$, $g$ has unique labels in $C$ if, and only if, $[g]$ has unique labels
  in $C_{\equiv}$. Indeed, for all $\sigma \in \sym_n$ if there exists $\pi \in
  \aut(C)$ extending $\sigma$ then $\pi /_\equiv$ is an automorphism of
  $C_\equiv$ extending $\sigma$.
  \label{lem:quotient-circuits-preserve}
\end{lem}
\begin{proof}
  Let $n$ be the order of $C$. We first prove that $C_\equiv$ and $C$ compute
  the same function. Let $\mathcal{A}$ be a $\rho$-structure of size $n$ and let
  $\gamma$ be a bijection from $A$ to $[n]$. We now show that for all $g \in G$,
  $C_\equiv [\gamma \mathcal{A}]([g]) = C[\gamma \mathcal{A}](g)$. We do this by
  induction on the depth of a gate. Suppose $g \in G$ has depth $0$. In this
  case $g$ is an input gate and the result follows trivially. Suppose $g$ is an
  internal gate and suppose for all $h$ of depth less than $g$ we have that
  $C_\equiv[\gamma \mathcal{A}]([h]) = C[\gamma \mathcal{A}](h)$. We have that
  there exists $g' \in [g]$ such that $L_\equiv ([g]) = L(g') /_\equiv$ and
  $\Sigma_\equiv([g]) = \Sigma (g) = \Sigma (g')$. We have from
  Lemma~\ref{lem:syntactic-equivalence-equal-function} and the inductive
  hypothesis that there exists $\lambda \in \natinv(g)$ such that $L_\equiv
  ([g]) = L(g') = L(g) \lambda$. It follows from the fact that $\Sigma(g)$ is a
  structured function that $C_\equiv[\gamma \mathcal{A}]([g]) = \Sigma_\equiv
  ([g])(L^{\gamma \mathcal{A}}_\equiv([g])) = \Sigma (g') (L^{\gamma
    \mathcal{A}}(g')) = \Sigma (g)(L^{\gamma \mathcal{A}}(g) \lambda) = \Sigma
  (g) (L^{\gamma \mathcal{A}}(g)) = C[\gamma \mathcal{A}](g)$.
  
  We now show that $C_\equiv$ is reduced. Suppose $[g], [h] \in G_\equiv$ and
  suppose $[g] \equiv [h]$. If $[g]$ and $[h]$ are both input or output gates
  then $[g] = [h]$. Suppose $[g]$ and $[h]$ are internal gates that are not
  output gates. Then $\Sigma (g) = \Sigma_\equiv ([g]) = \Sigma_\equiv([h]) =
  \Sigma(h)$. There exists $g', h' \in G$ such that $g' \equiv g$ and $h' \equiv
  h$, and $L_\equiv ([g]) = L(g') /_\equiv$ and $L_\equiv ([h]) = L(h')
  /_\equiv$. It follows from $[g] \equiv [h]$ that $L(g') /_\equiv$ is
  isomorphic to $L(h') /_\equiv$. From this it follows that $g' \equiv h'$, and
  so $[g] = [h]$.

  Let $\sigma \in \sym_n$ and suppose there exists $\pi \in \aut(C)$ extending
  $\sigma$. Let $\pi_\equiv = \pi /_\equiv$. We now show that $\pi_\equiv$ is an
  automorphism of $C_\equiv$ extending $\sigma$. It is easy to see that
  $\pi_\equiv$ is a bijection from $G_\equiv$ to $G_\equiv$ that preserves
  syntactic-equivalence, and is thus a well-defined function. Let $[g] \in
  G_\equiv$. We have that $\Sigma_\equiv (\pi_\equiv [g]) = \Sigma_\equiv ([\pi
  g]) = \Sigma (\pi g) = \Sigma (g) = \Sigma_\equiv ([g])$. It is easy to check
  the automorphism conditions for input gates. Suppose $[g]$ is an internal
  gate. Let $g' \in [g]$ be such that $L_\equiv([g]) = L(g') /_\equiv$ and $h
  \in [g]$ be such that $L_\equiv(\pi_\equiv[g]) = L_\equiv ([\pi h]) = L(\pi h)
  /_\equiv$. We have that $\pi L(g')$ is isomorphic to $L(\pi g')$, and it
  follows that $(\pi L(g'))/_\equiv$ is isomorphic to $L(\pi g') /_\equiv$. We
  then have $(\pi L(g'))/_\equiv = \pi_\equiv (L(g') /_\equiv) = \pi_\equiv
  L_\equiv ([g])$ and, since $g' \equiv h$ and so $\pi g' \equiv \pi h$, we have
  that $L(\pi g')/_\equiv$ is isomorphic to $L(\pi h) /_\equiv =
  L_\equiv(\pi_\equiv[g])$. It follows that $\pi_\equiv L_\equiv ([g])$ is
  isomorphic to $L_\equiv(\pi_\equiv [g]) /_\equiv$. Suppose $[g]$ is an output
  gate. Then for $\vec{a} \in \dom (\Omega)$, $\pi_\equiv \Omega_\equiv
  (\vec{a}) = [\pi \Omega (\vec{a})] = [\Omega (\sigma \vec{a})] = \Omega_\equiv
  (\sigma \vec{a})$. It follows that if $C$ is symmetric then $C_\equiv$ is
  symmetric.

  We now show that for $g \in G$, $g$ has unique labels in $C$ if, and only if,
  $[g]$ has unique labels in $C_{\equiv}$. Let $g \in G$ and let $h \in [g]$
  such that $L_\equiv([g]) = L(h) /_\equiv$. We first prove the backward
  direction. Suppose $[g]$ has unique labels. Then, since $L_\equiv([g])$ is
  injective, $L(h) /_\equiv$ must be injective and so $L(h)$ must be injective
  and no two child gates of $h$ can be syntactically-equivalent. It follows that
  $h$ has unique labels. Since $h \equiv g$ and $h$ has unique labels, it
  follows that $g$ has unique labels. We now prove the forward direction.
  Suppose $g$ has unique labels. Then, since $h \equiv g$ and $g$ has unique
  labels, $h$ has unique labels and so $L_\equiv([g])$ has injective labels.
  Since each syntactic equivalence class in $C_\equiv$ is a singleton it follows
  that $[g]$ has unique labels.
\end{proof}

We next show that there is a polynomial-time algorithm that takes as input a
transparent circuit and outputs a reduced injective circuit computing the same
function. Roughly, this algorithm first computes the syntactic equivalence
relation of the input circuit using Lemma~\ref{lem:transparent-syntactic-equiv}
and then, by picking representatives, defines a quotient circuit. It follows
from Lemma~\ref{lem:quotient-circuits-preserve} that the quotient circuit is
reduced and computes the same function as the input circuit, but it may have
fully symmetric gates that are not injective. To remedy this, we iterate over
the circuit, ``copying'' each child of each fully symmetric gate an appropriate
number of times and appending to each copy a distinct gadget that ensures that
no two copies can be syntactically equivalent. The complete proof of this
result, and the specific construction of these gadgets, is laborious but
straight forward. We omit the details here and refer to the reader
to~\cite{wilsenachthesis2019} for a complete proof.

\begin{lem}
  There is an algorithm that takes as input a transparent $(\mathbb{B},
  \rho)$-circuit $C$ and outputs a $(\mathbb{B} \cup \mathbb{B}_{\std},
  \rho)$-circuit $C'$ such that $C$ and $C'$ compute the same function, $C'$ is
  reduced and injective, and if $C$ is symmetric then $C'$ is symmetric.
  Moreover, this algorithm runs in time polynomial in the size of the input
  circuit.
  \label{lem:transparent-unique}
\end{lem}

We now show that there is a polynomial-time algorithm for computing the image of
a gate under the action of a permutation

\begin{lem}
  There is an algorithm that takes as input a reduced injective $(\BB,
  \rho)$-circuit $C$ of order $n$ and a permutation $\sigma \in \sym_n$ and
  outputs for each gate $g$ the image of $g$ under the action of the unique
  automorphism extending $\sigma$ (if it exists). This algorithm runs in time
  polynomial in the size of the input circuit.
  \label{lem:compute-automorphisms}
\end{lem}

\begin{proof}
  Let $\langle G, \Omega, \Sigma, \Lambda, L \rangle = C$. The algorithm we
  define works by inductively extending the action of $\sigma$ on the input
  gates to the rest of the circuit, or, if this extension fails, returning that
  there is no automorphism extending $\sigma$.

  Let $h$ be a gate in $C$. Suppose $h$ is an input gate. If $h$ is a constant
  gate then let $\pi (h) = h$. If $h$ is a relational gate check that there
  exists $h'$ such that, $\Sigma(h) = \Sigma(h')$, $\sigma \Lambda(h) =
  \Lambda(h')$, and if $h$ is an output gate then $h' = \Omega (\sigma
  (\Omega^{-1}(h)))$. If $h'$ exists let $\pi (h) = h'$, otherwise halt and
  output that no automorphism exists.

  Suppose that $h$ is an internal gate and we have defined $\pi$ for every gate
  of depth less than $h$. Check that there exists $h'$ such that $\Sigma(h) =
  \Sigma(h')$, $\pi L(h)$ is isomorphic to $L(h')$ and, if $h$ is an output gate
  then $h' = \Omega (\sigma (\Omega^{-1}(h)))$. If $h'$ exists let $\pi (h) =
  h'$, otherwise halt and output that no automorphism exists.

  If the induction continues until $\pi(g)$ has been defined for every gate in
  the circuit, halt and output $\pi$. Note that, since the circuit is reduced
  and injective, at each step in the inductive construction there is at most one
  $h'$ satisfying the relevant requirements and so $\pi$ is well defined. Note
  as well that $\pi$ is not, in general, an automorphism.
  
  We now show that the above algorithm outputs a function $\pi$ such that $\pi$
  is an automorphism extending $\sigma$ if, and only if, there exists an
  automorphism extending $\sigma$. The forward direction is trivial, so it
  suffices to show that if there exists an automorphism $\pi'$ extending
  $\sigma$ then the algorithm outputs a function $\pi$ and $\pi = \pi'$. We
  prove this by induction on depth. If $h$ is an input gate then we can verify
  that $\pi'(h)$ is the unique gate satisfying the above three requirements, and
  so $\pi(h) = \pi'(h)$. Suppose $h$ is an internal gate and $\pi(g) = \pi'(g)$
  for all gates $g$ of depth less than $h$. From the definition of an
  automorphism, $\Sigma(h) = \Sigma(\pi'(h))$, $\pi' L(h)$ is isomorphic to
  $L(\pi'(h))$, and if $h$ is an output gate then $\pi'(h) = \Omega (\sigma
  (\Omega^{-1}(h)))$. Since, from the induction hypothesis, $\pi' L(h)$ is
  isomorphic to $L(\pi'(h))$, we have that $\pi L(h)$ is isomorphic to
  $L(\pi'(h))$. Since $C$ is reduced and injective, $\pi'(h)$ is the unique gate
  satisfying these requirements, and so $\pi (h) = \pi'(h)$.

  It therefore suffices to run the above algorithm and if it outputs a function
  $\pi$ check that $\pi$ is an automorphism extending $\sigma$. If so output the
  image of each gate under $\pi$ and otherwise output that no automorphism
  exists.

  It remains to show that we can run the above algorithm and, if it outputs a
  function $\pi$, check if $\pi$ is an automorphism in time polynomial in the
  size of the circuit. It suffices to show that for any two gates $h$ and $h'$
  in $C$, we can check if $\pi L(h)$ is isomorphic to $L(h')$ in polynomial
  time. But, since this circuit is reduced and injective, $\pi L(h)$ is
  isomorphic to $L(h')$ if, and only if, $L^{-1}(h')  \pi L(h)$ is an
  automorphism. It can easily be checked if $L^{-1}(h')  \pi L(h)$ is
  an automorphism of the structure associated with $h$.

\end{proof}

We now use Lemma~\ref{lem:compute-automorphisms} to define an algorithm that
computes in polynomial time the image of a given element of the universe of a
gate under the action of a given permutation.

\begin{lem}
  There is an algorithm that takes as input an injective reduced $(\mathbb{B},
  \rho)$-circuit $C$ of order $n$, $\sigma \in \sym_n$, $g$ a gate in $C$, and
  $a \in \universe{g}$ and, if there exists an automorphism of $C$ extending
  $\sigma$ such that $\sigma \in \stab(g)$, outputs $\sigma(a)$. The algorithm
  runs in time polynomial in the size of $C$ and the encoding of $\sigma$.
  \label{lem:compute-automorphisms-labels}
\end{lem}
\begin{proof}
  Let $C = \langle G, \Omega, \Sigma, \Lambda, L \rangle$. We use the algorithm
  from Lemma \ref{lem:compute-automorphisms} to check if $\sigma$ extends to an
  automorphism on $C$. We also check if $\sigma \in \stab(g)$. If either of
  these checks fail, halt and return that no such automorphism exists. Let $h
  \in H_g$ and $\vec{b} := L(g)^{-1}(h)$ be such that $a \in \vec{b}$, and let
  $i \in [\vert \vec{b} \vert]$ be the index of $a$ in $\vec{b}$. Halt and
  output $\sigma a = (L(g)^{-1}(\sigma h))(i)$.
\end{proof}

We now show that there is a polynomial-time algorithm that takes as input a
reduced injective symmetric circuit $C$ and a gate $g$ and outputs the orbit and
canonical support of $g$.

\begin{lem}
  \label{lem:computing-support-orbit}
  There is an algorithm that takes as input an injective reduced symmetric
  circuit $C$ and a gate $g$ with a support of size less than $\frac{n}{2}$ and
  if $C$ is symmetric outputs $\orb(g)$ and $\consp(g)$ and otherwise outputs
  that $C$ is not symmetric. This algorithm runs in time polynomial in the size
  of the circuit.
\end{lem}
\begin{proof}
  Let $n$ be the order of $C$. Let $T \subseteq \sym_n$ be the set of
  transpositions. Note that $T$ generates $\sym_n$. We can thus determine if $C$
  is symmetric by running the algorithm in Lemma~\ref{lem:compute-automorphisms}
  for the gate $g$ and for each transposition in $T$. Since $\vert T \vert = {n
    \choose 2}$ the total time taken to execute this is polynomial in $\vert C
  \vert$. If $C$ is not symmetric we halt and output that $C$ is not symmetric.
  We assume for the rest of this proof that $C$ is symmetric.

  Let $\hat{T} = T \cap \stab(g)$. Let $\sim$ be the equivalence relation on
  $[n]$ defined such that $a \sim b$ if, and only if, $\trans{a}{b} \in
  \hat{T}$. To see that $\sim$ is an equivalence relation observe that if $a
  \sim b$ and $b \sim c$ then $\trans{a}{b}$ and $\trans{b}{c}$ are in $\hat{T}$
  and so, since $\trans{a}{c} = \trans{a}{b}\trans{b}{c}\trans{a}{b}$, we have
  $a \sim c$. Note that if $a, b \in [n] \setminus \consp(g)$ then $a \sim b$.
  It follows that there exists $S \in \quot{[n]}{\sim}$ such that $[n] \setminus
  \consp(g) \subseteq S$. Since $\vert [n] \setminus \consp(g) \vert >
  \frac{n}{2}$ we have that $S$ is the unique largest equivalence class in
  $[n]$.
  
  Suppose $S \neq [n] \setminus \consp(g)$. Since $[n] \setminus \consp(g)
  \subsetneq S$, there exists $c \in \consp(g)$ such that $[n] \setminus
  (\consp(g) \setminus \{c\}) \subseteq S$. Thus for every $a, b \in [n]
  \setminus (\consp(g) \setminus \{c\})$ we have $\trans{a}{b} \in \hat{T}$, and
  so $\trans{a}{b} \in \stab(g)$. But the set of all these transpositions
  generate $\stab(\consp(g) \setminus \{c\})$, and so $\stab(\consp(g) \setminus
  \{c\}) \leq \stab(g)$. This contradicts the minimality of $\consp(g)$, and so
  we conclude $S = [n] \setminus \consp(g)$, and so $\consp(g) = [n] \setminus
  S$.
\end{proof}

we can use similar methods to establish an analogous result for elements of the
universe of the gate. we omit the details of the proof but state the result
formally below.

\begin{lem}
  there is a polynomial-time algorithm that takes as input a transparent
  symmetric circuit $c$, a gate $g$ which has a support of size less than
  $\frac{n}{2}$, and an element $a \in \universe{g}$, and outputs
  $\consp_{g}(a)$ and $\orb_{g} (a)$.
  \label{lem:computing-support-orbit-index}
\end{lem}

\section{Translating Formulas into Circuits}\label{sec:formulas-to-circuits}

In order to prove one direction of our main result we need to establish a
translation from $\FPR$-formulas to $\PT$-uniform families of transparent
symmetric rank circuits. There is a known translation from $\FO$-formulas to
$\PT$-uniform families of symmetric circuits~\cite{immerman1999descriptive}.
This translation associates an $\FO$-formula $\theta(\vec{x})$ with a family of
circuits $(C_n)_{n \in \nats}$ such that each gate in $C_n$ corresponds to a
subformula $\psi(\vec{y})$ and an assignment $\vec{a}$ to $\vec{y}$ in $[n]$.
Each gate in $C_n$ is labelled according to the symbol at the head of the
associated subformula, with existential and universal quantifiers translating to
large disjunctions and conjunctions, respectively.

This approach can be extended to formulas of fixed-point logic by first
unfurling the fixed-point operator in order to define a family of $\FO$-formulas
(see~\cite{Kolaitis1992} for details) with a constant bound on width, and then
translating each of these formulas to a symmetric circuit. The bound on width is
ensured by renaming variables at each stage when unfurling the operator. This
approach can be extended to logics with counting, so long as we include
threshold or majority gates in the circuit and extend first-order logic with
counting quantifiers, yielding a translation from $\FPC$-formulas to bounded
width $\PT$-uniform families of symmetric majority circuits~\cite{AndersonD17}.

We should like to use an analogous approach to define a translation from $\FPR$
to bounded width $\PT$-uniform families of transparent rank circuits. The
problem is that these translations do not, in general, yield families of
\emph{transparent} circuits. The essential reason is that a formula might be
invariant under some non-trivial permutations of the bound variables, and the
translation to circuits described above preserves these ``syntactic
symmetries'', potentially resulting in gates that are not fully symmetric with
two syntactically equivalent children. We now sketch a translation that does
produce transparent circuits by adding some additional gadgetry to the formula
in order to remove symmetries of this sort. The full proof is available
here~\cite{wilsenachthesis2019}.

\begin{theorem}
  For each $\FPR$-formula $\theta(\vec{x})$ there exists a $\PT$-uniform family
  of transparent symmetric rank circuits $(C_n)_{n \in \nats}$ that defines the
  same query as $\theta(\vec{x})$.
  \label{thm:translating-formulas-to-circuits}
\end{theorem}

This theorem is proved in three steps. First, we introduce a logic $\FOrk$,
defined by extending $\FO$ with \emph{rank quantifiers}\footnote{Dawar et
  al.~\cite{Dawar09logicswith} define a different rank quantifier and a
  corresponding extension $\FO$. They establish a similar translation from a
  fixed-point logic with rank operators to families of first-order formulas with
  rank quantifiers. The translation is only defined for formulas in which the
  rank operator only binds element variables. We are interested in the case
  where both element and number variables may be bound by the operator, which
  requires introducing a quite different rank quantifier. The logic $\FOrk$ and
  the extension of first-order logic with rank quantifiers
  in~\cite{Dawar09logicswith} are thus defined quite differently and are not known
  to have the same expressive power.}. Second, we show that for each $\FPR$-formula
$\theta(\vec{x})$ there exists a $\PT$-uniform family of $\FOrk$-formulas
$(\theta_n(\vec{x}))_{n \in \nats}$ and $k \in \nats$, such that each
$\theta_n(\vec{x})$ defines the same query as $\theta(\vec{x})$ on structures of
size $n$ and has width bounded by $k$. Thirdly, we define an algorithm that
takes as input a natural number $n$, computes $\theta_n$ and then, by recursion
on the structure of the formula, outputs a corresponding transparent symmetric
rank circuit $C_n$ in time polynomial in $n$.

The rank quantifiers in $\FOrk$ determine if the rank of a matrix, defined using
a family of formulas, with entries taken to be in some fixed finite field is at
least equal to a given threshold. There are some subtleties involved in defining
these quantifiers formally, and we refer the reader
to~\cite{wilsenachthesis2019} for a complete definition of $\FOrk$. For our
purposes it suffices to note that $\FOrk$ is strong enough to define formulas of
the form
\begin{align*}
  \theta := \rank^{\geq t}_p [\vec{x}, \vec{y}] \phi(\vec{x}, \vec{y}),
\end{align*}
where $\theta$ holds for a given structure if the $0$-$1$ matrix defined by
$\phi$ with rows indexed by assignments to $\vec{x}$ and columns indexed by
assignments to $\vec{y}$ and interpreted as having entries in $\ff_p$ has rank
at least $t$.

The translation from $\FPR$-formulas to bounded width $\PT$-families of
$\FOrk$-formulas can be completed using essentially the same approach as
in~\cite{Kolaitis1992, Holm2010} for translating $\FPC$ formulas to uniform
families of $\FOc$-formulas. This works by unfurling the fixed-point operators,
i.e.\ recursively replacing each second-order variable bound by the operator
with the corresponding formula, and replacing number terms, while renaming bound
variables to keep the total number of variables bounded. When unfurling these
formulas we encode $\FOrk$-formula as a branching program so as to avoid
exponential blow-ups. The details of this translation are available
here~\cite{wilsenachthesis2019}.

The final step in this translation, the recursive construction of a transparent
symmetric circuit from an $\FOrk$-formula, also follows a similar approach to
the standard translation discussed above, but now with some additional gadgetry
added in order to remove symmetries in the formula. To understand what sorts of
symmetries are relevant and why the standard translation would result in
potentially non-transparent circuits consider as an example the $\FOrk$-formula
$\rank^{\geq t}_p [x, y] (E(x, x) \land E(y, y))$. The standard translation
would result in a rank gate with each child corresponding to a map from $\{x,
y\}$ to $[n]$. It is easy to see that for each $a, b \in \nats$ the child
corresponding to the assignment $(a, b) \mapsto (x, y)$ will be syntactically
equivalent to the child corresponding to the assignment $(b, a) \mapsto (x, y)$.
To remove these sorts of symmetries we preprocess each formula $\theta_n$ and
replace each subformula of the form $\rank^{\geq t}_p [\vec{x}, \vec{y}] \phi$
with one of the form $\rank^{\geq t}_p [\vec{x}, \vec{y}] (\phi \land \psi)$,
where $\psi$ is chosen so as to not be invariant under any permutation of
$\vec{x} \cup \vec{y}$.

\section{Translating Circuits into Formulas}\label{sec:circuits-to-formulas}

In this section we prove the following theorem establishing a translation
from uniform families of symmetric rank circuits to $\FPR$-formulas.

\begin{theorem}
  For each $\PT$-uniform family of transparent symmetric rank circuits $(C_n)_{n
    \in \nats}$ there exists an $\FPR$-formula expressing the same query.
  \label{thm:circuits-formulas}
\end{theorem}

The proof is structured as follows. Let $(C_n)_{n \in \nats}$ be a $\PT$-uniform
family of transparent symmetric rank circuits. From the Immerman-Vardi theorem
there exists a sequence of $\FP[\leq]$-formulas $\Phi$ such that $\Phi$ defines
$C_n$ when interpreted in the structure $([n], \leq)$. From
Lemma~\ref{lem:transparent-unique} we may assume, without loss of generality,
that each $C_n$ is reduced and injective. From
Theorem~\ref{thm:support-theorem-universe} there is a constant $k \in \nats$
bounding the size of the supports of each gate and each element of the universe
of a gate in each $C_n$. We show in
Lemma~\ref{lem:support-determines-evaluation} that for a given $n$-size
structure $\mathcal{A}$ and bijection $\gamma : A \ra [n]$ the evaluation of a
gate $g$ in $C_n$ depends only on how $\gamma$ maps elements of $\mathcal{A}$ to $\consp(g)$. It follows that each gate $g$ in some $C_n$ can be
associated with a fixed arity relation $\EV_g$ consisting of all assignments to
$\consp(g)$ for which $g$ evaluates to $1$. We use the various algorithms
discussed in Section~\ref{sec:algorithms} and the Immerman-Vardi theorem to show that we can
recursively define each $\EV_g$ in $\FPR$, allowing us to evaluate the output
gates in $\FPR$ and so prove the result.

This overall approach is similar in structure to the one used by Anderson and
Dawar~\cite{AndersonD17} for translating $\PT$-uniform families of symmetric majority circuits
to $\FPC$-formulas. However, the important results so far discussed, e.g.\ the
support theorem and the existence of polynomial-time algorithms computing
circuit properties, do not generalise obviously and have required quite
different arguments. The work of this section draws these results together and
establishes the final step in the translation, the inductive definition of
$\EV_g$ for each gate $g$ in the circuit. Anderson and Dawar establish this by
first exhibiting a bijection from the orbit of a gate to the set of assignments
to its support and then, using this bijection, counting the number of children
of $g$ that evaluate to $1$. This suffices as the gates in the circuits of
interest there are all fully symmetric, and so counting the
number of inputs that evaluate to $1$ suffices to evaluate the gate. However, in
our more general setting a gate could be labelled by a rank function. The
evaluation of such a gate depends not just on the \emph{number} of children that
evaluate to true, but on the \emph{structure} defined at $g$.

The majority of the work of this section is in showing that this structure can
be recursively defined in $\FPR$. In
Section~\ref{sec:recursive-matrix} we give a recursive construction of a
structure and show
that it suffices. In Section~\ref{sec:translating-formulas-to-FPR} we show how
this construction can
be implemented in $\FPR$ and complete the proof
of Theorem~\ref{thm:circuits-formulas}.

\subsection{Recursively Defining a Matrix}
\label{sec:recursive-matrix}
We first show that for a gate $g$ in a symmetric circuit $C$ of order $n$ the
evaluation of $g$ for an input structure $\mathcal{A}$ and bijection $\gamma \in
[n]^{\underline{A}}$ depends only how $\gamma$ maps to $\consp(g)$. We say two
injective functions $f$ and $g$ are \emph{compatible} if we can define an
injection on the union of their domains that agrees with each function on their
respective domains. We write $f \sim g$ to denote that $f$ and $g$ are
compatible and $f \vert g$ to denote the injection defined on the union of their
domains and that agrees with each function on their respective domains.

\begin{lem}
  Let $C$ be a symmetric rank circuit with vocabulary $\rho$ of order $n$. Let
  $\mathcal{A} \in \fin{\rho, n}$. Let $g$ be an internal gate, $\eta \in
  A^{\underline{\consp(g)}}$, and $\gamma_1, \gamma_2 \in [n]^{\underline{A}}$
  such that $\gamma^{-1}_1 \sim \eta$ and $\gamma^{-1}_2 \sim \eta$. Then
  $L^{\gamma_1 \mathcal{A}}(g)$ and $L^{\gamma_2 \mathcal{A}}(g)$ are isomorphic
  and consequently $C[\gamma_1 \mathcal{A}](g) = C[\gamma_2 \mathcal{A}](g)$.
	\label{lem:support-determines-evaluation}
\end{lem}
\begin{proof}
	Let $\pi \in \sym_n$ be the unique permutation such that $\pi \gamma_1 =
  \gamma_2$. Since $\gamma^{-1}_1$ and $\gamma^{-1}_2$ are both compatible with
  $\eta$, it follows that $\pi$ must fix $\consp(g)$ pointwise. From the
  definition of a support, we have that $\pi (g) = g$, and so $L(g)$ is
  isomorphic to $\pi L(g)$. Therefore there exists $\lambda \in \natinv(g)$ such
  that $\pi L(g) = L(g) \lambda$, and so for all $a \in \ind(g)$,
	\begin{align*}
		L^{\gamma_1 \mathcal{A}}(g) (a) & = C [\gamma_1 \mathcal{A}](L(g)(a))                    \\
                                    & = C[\pi \gamma_1 \mathcal{A}][\pi L(g)(a)]            \\
                                    & = C[\gamma_2 \mathcal{A}][L(g)(\lambda(a))] \\
                                    & = L^{\gamma_2 \mathcal{A}}(g) (\lambda (a)).                 
	\end{align*}
	It follows that $L^{\gamma_1 \mathcal{A}}(g)$ and $L^{\gamma_2
    \mathcal{A}}(g)$ are isomorphic and so $C[\gamma_1 \mathcal{A}](g) =
  \Sigma(g) (L^{\gamma_1 \mathcal{A}}(g)) = \Sigma(g) (L^{\gamma_2
    \mathcal{A}}(g)) = C[\gamma_2 \mathcal{A}](g)$.
\end{proof}

For the remainder of this subsection we fix a $\PT$-uniform family of
transparent symmetric rank circuits $\mathcal{C} = (C_n)_{n \in \nats}$ over a
vocabulary $\rho$. By Corollary~\ref{cor:poly-size-support-bound} there are
constants $n_0$ and $k$ such that for all $n > n_0$, we have $\SPs{C_n} \leq k$.
Fix some $n > \max(n_0, 3k)$ and $\mathcal{A} \in \fin{\rho, n}$. Let $C :=
\langle G, \Omega, \Sigma, \Lambda, L \rangle := C_n$.

For each $g \in G$ let $\Gamma_g:= \{\gamma \in [n]^{\underline{A}} : C[\gamma
\mathcal{A}](g) = 1 \}$ and let $\EV_g := \{ \eta \in A^{\underline{\consp(g)}}
: \exists \gamma \in \Gamma_g \, ,\eta \sim \gamma^{-1}\}$. In other words,
$\Gamma_g$ is the set of bijections for which $g$ evaluates to $1$ and $\EV_g$
is the set of assignments to the support of $g$ that can be extended to a
bijection for which $g$ evaluates to $1$. It follows from
Lemma~\ref{lem:support-determines-evaluation} that $\Gamma_g$ is entirely
determined by $\EV_g$. From Theorem~\ref{thm:support-theorem-universe} the
domain of each $\eta \in \EV_g$ has cardinality at most $k$. In this sense
$\EV_g$ can be thought of as a succinct encoding of $\Gamma_g$.

We aim to define the set $\EV_g$ for each gate $g$ in $C$ by induction on the
structure of the circuit. This definition depends on the basis function
labelling $g$ and has already been given in~\cite{AndersonD17} for gates
labelled by $\AND$, $\OR$, $\NOT$, or $\MAJ$ functions. We therefore restrict
our attention to the case where $g$ is lablelled by a $\RANK$ function.

For the remainder of this subsection we fix a gate $g \in G$ labelled by a rank
function $\RANK^t_p[X]$ with $X = \uplus_{s \in [3]}X_s$. We also fix some $\eta
\in A^{\underline{\consp(g)}}$.

We now introduce some notation. For $\gamma \in [n]^{\underline{A}}$ we write
$L^{\gamma}$ to abbreviate $L^{\gamma \mathcal{A}}(g)$. For each $x \in
\universe{g}$ we abbreviate notation and let $\consp(x) := \consp_{g}(x)$,
$\orb(x) := \orb_{g}(x)$, and $\stab(x) := \stab_{\spstab{g}}(x)$. For each $h
\in H_g$ let $A^h := \{\alpha \in A^{\underline{\consp(h)}} : \eta \sim
\alpha\}$ be the set of assignments to the support of $h$ that are compatible
with $\eta$. For each $x \in \universe{g}$ let $A^x:= \{\alpha \in
A^{\underline{\consp(x)}} : \eta \sim \alpha\}$ be the set of assignments to the
support of $x$ that are compatible with $\eta$.

In Section~\ref{sec:translating-formulas-to-FPR} we encode each circuit as a
structure over a linearly ordered universe, which then induces an order on the
universe of each gate. As such, in this section we suppose for each $s \in [3]$
that there is some linear order on $X_s$. This allows us to choose a
representative from each orbit. So, for each $s \in [3]$ let $\minorb{s} :=
\{\min (\orb(x)) : x \in X_s\}$ and let $I^{g, \eta}_{s} := \{(x, \alpha): x \in
\minorb {s}, \alpha \in A^x\}$. Let $I^{g, \eta} := \uplus_{s \in [3]}I^{g,
  \eta}_s$. For each $s \in [3]$ we think of each $x \in \minorb{s}$ as denoting
an orbit in $X_s$ and each assignment in $A^x$ as encoding the action of a
permutation in $\stab(g)$ on $\consp(x)$, which determines an element of that
orbit. In this way we think of each element in $I^{g, \eta}_s$ as encoding an
element in $X_s$.

We next define a three sorted matrix $M^{g, \eta}: \tot{\matvoc}{I^{g, \eta}}
\ra \{0, 1\}$ and show that $\rank_p (M^{g, \eta}) = \rank_p(L^{\gamma})$. We do
so by first defining a function $J^{g, \eta}$ that maps each $((x_1, \alpha_1),
(x_2, \alpha_2) (x_3, \alpha_3)) \in \tot{\matvoc}{I^{g, \eta}}$ to a triple
$(\sigma_1, \sigma_2, \sigma_3)$ of permutations in $\stab(g)$ with the property
that the assignments defined by taking the action of each $\sigma_i$ on
$\alpha_i$ are pairwise compatible. We let $\epsilon$ be the injection formed by
combining these assignments and $h := L(g)(\sigma_1(x_1), \sigma_2(x_2),
\sigma_3(x_3))$. We then define $M^{g, \eta}$ by letting $M^{g, \eta}((x_1,
\alpha_1), (x_2, \alpha_2), (x_3, \alpha_3)) = 1$ if, and only if, $\epsilon \in
\EV_h$. We now complete this definition formally.

As an intermediary step we first define a function $\bar{J}^{g, \eta}$ that maps
each $((x_1, \alpha_1), (x_2, \alpha_2) (x_3, \alpha_3)) \in \tot{\matvoc}{I^{g,
    \eta}}$ to a sequence of injections $(\bar{\sigma}_1, \bar{\sigma}_2,
\bar{\sigma}_3)$ defined as follows. Let $B \subseteq [n] \setminus \consp(g)$
have size $3k$ and let $B := B_1 \cup B_2 \cup B_3$ be a partition of $B$. For
each $i \in [3]$ let $f_i : \dom(\alpha_i) \setminus \consp(g) \ra B_i$ be an
injection.
\begin{itemize}
\item Let $\bar{\sigma}_1 : \dom(\alpha_1) \ra [n]$ be defined for $u \in
  \dom(\alpha_1)$ such that if $u \in \consp(g)$ then $\bar{\sigma}_1(u) = u$
  and otherwise $\bar{\sigma}_1 (u) = f_1(u)$.
\item Let $\bar{\sigma}_2 : \dom(\alpha_2) \ra [n]$ be defined for $u \in
  \dom(\alpha_2)$ such that if $u \in \dom(\alpha_1)$ then $\bar{\sigma}_2(u) =
  \bar{\sigma}_1(u)$ and otherwise $\bar{\sigma}_2(u) = f_2 (u)$.
\item Let $\bar{\sigma}_3 : \dom(\alpha_3) \ra [n]$ be defined for $u \in
  \dom(\alpha_3)$ such that if $u \in \dom (\alpha_1)$ then $\bar{\sigma}_3(u) =
  \bar{\sigma}_1(u)$ and if $u \in \dom(\alpha_2)$ then $\bar{\sigma}_3(u) =
  \bar{\sigma}_2(u)$ and otherwise $\bar{\sigma}_3(u) = f_3(u)$.
\end{itemize}

We now make a few observations. We first note that each $\bar{\sigma}_i$ is an
injection. We abuse notation slightly and write $\alpha_i \circ
\bar{\sigma}^{-1}_i$ to denote the injection $\alpha_i \circ \bar{\sigma}^{-1}_i
: \img(\bar{\sigma}_i) \ra A$ defined by $\alpha_i \circ \bar{\sigma}^{-1}_i (u)
= \alpha_i(\bar{\sigma}^{-1}_i (u))$ for all $u \in \img(\bar{\sigma}_i)$. We
think of each $\alpha_i \circ \bar{\sigma}^{-1}_i$ as being a version of
$\alpha_i$ with the domain shifted by $\bar{\sigma}_i$. It can be shown that the
three functions $\alpha_i \circ \bar{\sigma}^{-1}_i $ for $i \in [3]$ are
pairwise compatible. Lastly, note that $\consp(g) \subseteq
\dom(\bar{\sigma}_i)$ and for each $u \in \consp(g)$ we have $\bar{\sigma}_i (u)
= u$. We summarise these assertions in the following lemma.

\begin{lem}
  Let $z := ((x_1, \alpha_1), (x_2, \alpha_2) (x_3, \alpha_3)) \in \matvoc{I^{g,
      \nu}}$ and let $(\bar{\sigma}_1, \bar{\sigma}_2, \bar{\sigma}_3) :=
  \bar{J}^{g, \eta}(z)$. Then for all $j \in [3]$
  \begin{myenum}
  \item for all $u \in \consp(g)$, $\bar{\sigma}_j (u) = u$,
  \item $\bar{\sigma}_j$ is an injection, and
  \item for all $j' \in [3]$, $\alpha_j \circ \bar{\sigma}^{-1}_j$ and
    $\alpha_{j'} \circ \bar{\sigma}^{-1}_{j'}$ are compatible.
  \end{myenum}
  \label{lem:bar-J-construction}
\end{lem}

Let $J^{g, \eta} : \tot{\matvoc}{I^{g, \eta}} \ra \{ (h, \epsilon) : h \in H_g,
\epsilon \in A^h\}$ be defined for $z := ((x_1, \alpha_1), (x_2, \alpha_2),
(x_3, \alpha_3)) \in \tot{\matvoc}{I^{g, \eta}}$ as follows. Let
$(\bar{\sigma}_1, \bar{\sigma}_2, \bar{\sigma}_3) := \bar{J}^{g, \eta}(z)$. For
each $i \in [3]$ let $\sigma_i \in \spstab{g}$ be such that for all $u \in
\consp(x_i)$, $\sigma_i (u) = \bar{\sigma}_i(u)$. Let $h := L(g)((\sigma_1
(x_1), \sigma_2 (x_2), \sigma_3 (x_3))$ and let $\epsilon := \restr{(\alpha_1
  \circ \bar{\sigma}^{-1}_1 \vert \alpha_2 \circ \bar{\sigma}^{-1}_2 \vert
  \alpha_3 \circ \bar{\sigma}^{-1}_3)}{\consp(h)}$. It follows from
Lemma~\ref{lem:bar-J-construction} that $\epsilon$ is well defined. Let $J^{g,
  \eta} (z) := (h, \epsilon)$.

Let $M^{g, \eta} : \tot{\matvoc}{I^{g, \eta}} \ra \{0,1\}$ be defined for $z:=
((x_1, \alpha_1), (x_2, \alpha_2), (x_3, \alpha_3)) \in \vatoms{\tau}{I^{g,
    \eta}}$ such that if $(h, \epsilon) := J^{g, \eta}(z)$ then $M^{g, \eta}(z)
= 1$ if, and only if, $\epsilon \in \EV_h$.

We stated previously that we can think of each assignment to the support of a
child or element of the universe of $g$ as encoding an element in its orbit.. We
now formalise this and show that for each $\gamma \in [n]^{\underline{A}}$ with
$\gamma^{-1} \sim \eta$ and each $h \in H_g$ there is a natural surjective map
defined by $\gamma$ from $A^h$ to $\orb(h)$ and for each $x \in \universe{g}$
there is a similarly defined surjection from $A^x$ to $\orb(x)$. Let $\gamma \in
[n]^{\underline{A}}$ be such that $\gamma^{-1} \sim \eta$. Let $h \in H_g$. For
each $\epsilon \in A^{h}$ let $\Pi^{\gamma}_{\epsilon}$ be any permutation in
$\spstab{g}$ such that $\Pi^{\gamma}_\epsilon (u) = \gamma (\epsilon(u))$ for
all $u \in \consp(h)$. This action of $\Pi^{\gamma}_\epsilon$ on $h$ is defined
independently of the choice of permutation, and so the function $\epsilon
\mapsto \Pi^{\gamma}_{\epsilon}(h)$ is well defined. Let $x \in X$. For each
$\alpha \in A^{x}$ let $\Pi^{\gamma}_{\alpha}$ be any permutation in
$\spstab{g}$ such that $\Pi^{\gamma}_\alpha (u) = \gamma (\alpha(u))$ for all $u
\in \consp(x)$. It follows similarly that the action of $\Pi^{\gamma}_\alpha $
on $x$ is defined independently of the choice of permutation and so the function
$\alpha \mapsto \Pi^{\gamma}_{\alpha} (x)$ is well defined. It is easy to show
that both of these maps are surjective.

We now define a function from $M^{g, \eta}$ to $L^{\gamma}$ (as
$\matvoc$-structures). Let $\gamma \in [n]^{\underline{A}}$ be such that
$\gamma^{-1} \sim \eta$. For each $s \in [3]$ let $P^{\gamma}_s: I^{g, \eta}_s
\rightarrow X_s$ be defined by $P^{\gamma}_s (x, \alpha) :=
\Pi^{\gamma}_{\alpha}(x)$ for all $(x, \alpha) \in I^{g, \eta}_s$. Let
$P^{\gamma} = \uplus_{s \in \bm{S}} P^{\gamma}_s$. We aim to show that
$P^{\gamma}$ defines an epimorphism from $M^{g, \eta}$ to $L^{\gamma
  \mathcal{A}}(g)$. We first establish a correspondence between $\EV_h$ and
those elements of the orbit of $h$ that evaluate to $1$.

\begin{lem}
	Let $\gamma\in [n]^{\underline{A}}$ be such that $\gamma^{-1} \sim \eta$ and
  $h \in H_g$. Then for all $\epsilon \in A^h$, $\epsilon \in \EV_h$ if, and
  only if, $C[\gamma \mathcal{A}](\Pi^{\gamma}_{\epsilon} (h)) = 1$.
  \label{lem:translate-EV-circuits}
\end{lem}
\begin{proof}
  From the definition of $\EV_h$ it follows that $\epsilon \in \EV_h$ if, and
  only if, there exists $\delta \in [n]^{\underline{A}}$ such that $\delta^{-1}
  \sim \eta$, $\delta^{-1} \sim \epsilon$, and $C[\delta \mathcal{A}](h) = 1$.
  Let $\pi := \gamma \circ \delta^{-1}$. Then $\pi \in \spstab{g}$ and $\pi
  \circ \delta = \gamma$. For $a \in \consp(h)$ we have $\gamma (\epsilon(a)) =
  \pi (\delta (\epsilon (a))) = \pi(a)$. The second equality follows from the
  fact that $\delta^{-1} \sim \epsilon$ and so $\delta (\epsilon (a)) = a$. Then
  for all $a \in \consp(h)$ we have $\Pi^{\gamma}_\epsilon (a) = \gamma
  (\epsilon (a)) = \pi(a)$, and so $\Pi^{\gamma}_{\epsilon}(h) = \pi(h)$. Thus
  \begin{align*}
    \epsilon \in \EV_h \iff C[\delta \mathcal{A}](h) = 1 
    & \iff C[\pi \circ \delta \mathcal{A}](\pi (h)) = 1 \\
    & \iff C[\gamma \mathcal{A}](\pi (h)) = 1 
      \iff C[\gamma \mathcal{A}](\Pi^{\gamma}_{\epsilon}(h)) = 1.
  \end{align*}
\end{proof}

The function $J^{g, \eta}$ maps triples of the form $((x_1, \alpha_1), (x_2,
\alpha_2), (x_3, \alpha_3))$ to pairs of the form $(h, \epsilon)$. We now show
that $\Pi^{\gamma}_{\epsilon}(h)$ is exactly the gate defined by mapping each
$x_i$ to $\Pi^{\gamma}_{\alpha_i}(x_i)$.

\begin{lem}
  Let $\gamma \in [n]^{\underline{A}}$ be such that $\gamma^{-1} \sim \eta$. Let
  $((x_1, \alpha_1), (x_2, \alpha_2), (x_3, \alpha_3)) \in \ind(M^{g, \eta})$
  and let $(h, \epsilon) := J^{g, \eta}(((x_1, \alpha_1), (x_2, \alpha_2), (x_3,
  \alpha_3)))$. Then $\Pi^{\gamma}_{\epsilon} (h) = L(g)(\Pi^{\gamma}_{\alpha_1}
  (x_1),\Pi^{\gamma}_{\alpha_2} (x_2), \Pi^{\gamma}_{\alpha_3} (x_3))$
  \label{lem:defining-h-from-index}
\end{lem}
\begin{proof}
  Let $\sigma_1, \sigma_2, \sigma_3 \in \spstab{g}$ be the permutations
  introduced in the definition of $J^{g, \eta}$. Let $i \in [3]$ and let $u \in
  \consp(x_{i})$. Then $\sigma_i (u) \in \consp(\sigma (x_i)) \subseteq
  \consp(h) \cup \consp(g)$. Suppose $\sigma_i (u) \in \consp(g)$. Then since
  $\sigma_i \in \spstab{g}$ we have $\Pi^{\gamma}_{\epsilon} (\sigma_i (u)) = u
  = \Pi^{\gamma}_{\alpha_i}(u)$. Suppose instead that $\sigma_i(u) \in
  \consp(h)$. Then $\Pi^{\gamma}_{\epsilon}(\sigma_i (u)) = \gamma (\epsilon
  (\sigma_i(u))) = \gamma (\alpha_i (\bar{\sigma}^{-1}_i (\sigma_i (u)))) =
  \gamma(\alpha_i (u)) = \Pi^{\gamma}_{\alpha_i} (u)$. In either case we can
  conclude that $\Pi^{\gamma}_{\epsilon} (\sigma_i(x_j)) =
  \Pi^{\gamma}_{\alpha_i}(x_i)$, and so
  \begin{align*}
    \Pi^{\gamma}_{\epsilon} (h) &= \Pi^{\gamma}_{\epsilon} (L(g)(\sigma_1 (x_1), \sigma_2 (x_2), \sigma_3 (x_3))  \\ &= L(g)(\Pi^{\gamma}_{\epsilon}(\sigma_1 (x_1)), \Pi^{\gamma}_{\epsilon}(\sigma_2 (x_2)), \Pi^{\gamma}_{\epsilon}(\sigma_3 (x_3))) \\ &= L(g)(\Pi^{\gamma}_{\alpha_1}(x_1), \Pi^{\gamma}_{\alpha_2}(x_2), \Pi^{\gamma}_{\alpha_3}(x_3))
  \end{align*}
\end{proof}

We now show that $P^{\gamma}$ is a homomorphism from $M^{g, \eta}$ to
$L^{\gamma}$.

\begin{lem}
  Let $((x_1, \alpha_1), (x_2, \alpha_2), (x_3, \alpha_3)) \in
  \tot{\matvoc}{I^{g, \eta}}$. Let $\gamma \in [n]^{\underline{A}}$ be such that
  $ \gamma^{-1} \sim \eta$. Then $M^{g, \eta}(((x_1, \alpha_1), (x_2, \alpha_2),
  (x_3, \alpha_3))) = L^{\gamma}(P^{\gamma}(x_1, \alpha_1), P^{\gamma}(x_2,
  \alpha_2), P^{\gamma}(x_3, \alpha_3))$.
	\label{lem:M-to-L-homomorphism}
\end{lem}
\begin{proof}
  Let $(h, \epsilon) := J^{g, \eta}(z)$. Then
  \begin{align*}
    M^{g, \eta}(z) = 1 &\iff \epsilon \in \EV_h  \\ &\iff C[\gamma \mathcal{A}] (\Pi^{\gamma}_{\epsilon}(h)) = 1 \\ &\iff C[\gamma \mathcal{A}] (L(g)(\Pi^{\gamma}_{\alpha_1} (x_1), \Pi^{\gamma}_{\alpha_2} (x_2), \Pi^{\gamma}_{\alpha_3} (x_3)) = 1 \\ &\iff L^{\gamma}(P^{\gamma}(x_1, \alpha_1),P^{\gamma}(x_2, \alpha_2), P^{\gamma}(x_3, \alpha_3)) = 1.
  \end{align*}
  The second equivalence follows from Lemma~\ref{lem:translate-EV-circuits} and
  the third equivalence follows from Lemma~\ref{lem:defining-h-from-index}.
\end{proof}

\begin{lem}
	If $\gamma \in [n]^{\underline{A}}$ is such that $\gamma^{-1} \sim \eta$ then
  $P^{\gamma}$ is surjective.
  \label{lem:M-to-L-surjective}
\end{lem}
\begin{proof}
  Let $x \in X$. Let $y = \min(\orb(x))$. There exists $\sigma \in \spstab{g}$
  such that $\sigma (y) = x$. Let $\alpha := \gamma^{-1} \circ
  \restr{\sigma}{\consp(y)}$. Then for each $u \in \consp(y) \cap \consp(g)$ we
  have $\alpha(u) = \gamma^{-1} \circ \restr{\sigma}{\consp(y)} (u) =
  \gamma^{-1}(u) = \eta (u)$, and so $\alpha \in A^y$. For each $u \in
  \consp(y)$ we have $\Pi^{\gamma}_\alpha(u) = \gamma (\alpha(u)) =
  \gamma(\gamma^{-1} \restr{\sigma}{\consp(y)}(u)) = \sigma (u)$. It follows
  that $P^{\gamma}(y, \alpha) = \Pi^{\gamma}_\alpha(y) = \sigma (y) = x$.
\end{proof}

We have from Lemmas~\ref{lem:M-to-L-homomorphism}
and~\ref{lem:M-to-L-surjective} that $P^\gamma$ defines an epimorphism from
$M^{g, \eta}$ to $L^\gamma$. But $P^{\gamma}$ is not in general an isomorphism.
The following result provides a useful characterisation of preimages under
$P^\gamma$.

\begin{lem}
  For all $(x, \alpha), (y, \beta) \in I^{g, \eta}_3$, $P(x, \alpha) = P(y,
  \beta)$ if, and only if, $x = y$ and there exists $\pi \in \stab(x)$ such that
  $\alpha (u) = \beta \circ \pi (u)$ for all $u \in \consp(x)$.
  \label{lem:mut-stab}
\end{lem}
\begin{proof}
  $\Rightarrow$: Suppose $P(x, \alpha) = P(y, \beta)$. Let $\sigma :=
  (\Pi^\gamma_{\alpha})^{-1} \circ \Pi^\gamma_{\beta}$. Then $\sigma (x) =
  (\Pi^\gamma_{\beta})^{-1} \circ \Pi^\gamma_{\alpha} (x) = y$. But then $x$ and
  $y$ are in the same orbit and are both minimum elements of that orbit. It
  follows that $x = y$ and so $\sigma \in \stab(x)$. Moreover, for all $u \in
  \consp(x)$, $\gamma(\alpha(u)) = \Pi^\gamma_\alpha(u) =
  \Pi^\gamma_\beta(\sigma (u)) = \gamma(\beta(\sigma(u)))$. Since $\gamma$ is a
  bijection it follows that $\alpha(u) = \beta \circ \sigma (u)$ for all $u \in
  \consp(x)$.

  $\Leftarrow$: Let $\sigma \in \stab(x)$ such that $\alpha (u) = \beta \circ
  \sigma (u)$ for all $u \in \consp(x)$. Then $\Pi^{\gamma}_\beta (\sigma(u)) =
  \gamma (\beta (\sigma(u))) = \gamma (\alpha (u)) = \Pi^{\gamma}_\alpha(u)$ and
  so $\Pi^{\gamma}_\beta (x) = \Pi^{\gamma}_\beta (\sigma (x)) =
  \Pi^{\gamma}_\alpha(x) = \Pi^{\gamma}_\alpha(y)$.
\end{proof}

We could informally think of $M^{g, \eta}$ as containing numerous copies of the
entries in $L^\gamma$. On this basis it might be conjectured then that the ranks
of the associated matrices are equal, as repeated rows and columns in $M^{g,
  \eta}$ would have no effect on the rank. This is not the case as the matrix
associated with $M^{g, \eta}$ is defined by summing over the third sort, and so
repeated entries effect the values in the matrix. We now define a new matrix
from $M^{g, \eta}$ that corrects for these duplicates.

Let $M^{g, \eta}_{*} : I^{g, \eta}_1 \times I^{g, \eta}_2 \ra \natz$ be defined
for $(x_1, \alpha_1), (x_2, \alpha_2) \in I^{g, \eta}_1 \times I^{g, \eta}_2$
such that
\begin{align}
  \label{eq:M-star}
  M^{g, \eta}_{*} ((x_1, \alpha_1), (x_2, \alpha_2)) = \sum_{(x_3, \alpha_3) \in I^{g, \eta}_3} \frac{M^{g, \eta}((x_1, \alpha_1), (x_2, \alpha_2), (x_3, \alpha_3))}{\vert \orb_{\stab(x_3)}(\alpha_3) \vert}.
\end{align}
Let $L^\gamma_* : X_1 \times X_2 \ra \natz$ be the matrix associated with
$L^{\gamma}$. We now show that $P^{\gamma}$ is a homomorphic mapping from $M^{g,
  \eta}_{*}$ to $L^{\gamma}_*$.
\begin{lemma}
  Let $\gamma \in [n]^{\underline{A}}$ be such that $\gamma^{-1} \sim \eta$.
  Then for all $((x_1, \alpha_1), (x_2, \alpha_2)) \in I^{g, \eta}_1 \times
  I^{g, \eta}_2$ we have $M^{g, \eta}_{*} ((x_1, \alpha_1), (x_2, \alpha_2)) =
  L^{\gamma}_*(P^{\gamma}(x_1, \alpha_1), P^{\gamma}(x_2, \alpha_2))$.
  \label{lem:sum-equiv}
\end{lemma}
\begin{proof}
  Let $((x_1, \alpha_1), (x_2, \alpha_2)) \in I^{g, \eta}_1 \times I^{g,
    \eta}_2$. Then
  \begin{align*}
    L^{\gamma}_*(P^{\gamma}(x_1, \alpha_1), P^{\gamma}(x_2, \alpha_2)) &= \vert \{w \in X_3 : L^\gamma(P^{\gamma}(x_1, \alpha_1), P^{\gamma}(x_2, \alpha_2), w) = 1\}\vert \\
                                                                       &= \sum_{w \in X_3} L^\gamma(P^{\gamma}(x_1, \alpha_1), P^{\gamma}(x_2, \alpha_2), w)  \\
                                                                       &= \sum_{(x_3, \alpha_3) \in I^{g, \eta}_3} \frac{L^\gamma(P^{\gamma}(x_1, \alpha_1), P^{\gamma}(x_2, \alpha_2), P^{\gamma}(x_3, \alpha_3))}{\vert (P^{\gamma})^{-1}[P^{\gamma}(x_3, \alpha_3)]\vert} \\
                                                                       &= \sum_{(x_3, \alpha_3) \in I^{g, \eta}_3} \frac{M^{g, \eta}((x_1, \alpha_1), (x_2, \alpha_2),(x_3, \alpha_3))}{\vert (P^{\gamma})^{-1}[P^{\gamma}(x_3, \alpha_3)]\vert}
  \end{align*}
  The third equivalence follows from Lemma~\ref{lem:M-to-L-surjective} and the
  fourth equivalence follows from Lemma~\ref{lem:M-to-L-homomorphism}. It
  follows from Lemma~\ref{lem:mut-stab} that
  \begin{align*}
    \vert (P^{\gamma})^{-1}[P^{\gamma}(x_3, \alpha_3)]\vert &= \vert \{(z, \delta) : P^\gamma (z, \delta ) = P^{\gamma}(x_3, \alpha_3)\}  \vert \\
                                                            &= \vert \{(z, \delta) : z = x_3 \land \exists \sigma \in \stab(x_3) \forall u \in \consp(x) \alpha_3 (u) = \delta \circ \sigma(u)\} \vert \\
                                                            &= \vert \orb_{\stab(x_3)}(\alpha_3) \vert,
  \end{align*} from which the result follows.
\end{proof}

We now show that for any prime $p$, $\rank_p(M^{g, \eta}_*) =
\rank(L^\gamma_*)$. We recall that the rank of a matrix is equal to the maximal
order of a non-zero minor. We prove this result by establishing a correspondence
between non-zero minors of these two matrices.

\begin{thm}
  Let $\gamma \in A^{\underline{n}}$ be such that $\gamma^{-1} \sim \eta$ and
  let $p \in \nats$ be prime. Then $\rank_p (M^{g, \eta}_*) = \rank_p
  (L^{\gamma}_*)$.
  \label{thm:rank-equivilence}
\end{thm}
\begin{proof}
  It follows from Lemmas~\ref{lem:M-to-L-homomorphism}
  and~\ref{lem:M-to-L-surjective} that $L^{\gamma}_*$ appears as a submatrix of
  $M^{g, \eta}_*$. Then for each $r \in \natz$ if $L^{\gamma}_*$ has a non-zero
  $r \times r$ minor then so does $M^{g, \eta}_*$ and so $\rank_p(L^{\gamma}_*)
  \leq \rank_p (M^{g, \eta}_*)$.

  Let $r = \rank_p(M^{g, \eta}_*)$ and let $S$ be an $r \times r$ submatrix of
  $M^{g, \eta}_*$ with non-zero determinant. Let $R \subset I^{g, \eta}_1$ and
  $C \subset I^{g, \eta}_2$ be sets indexing the rows and columns of $S$,
  respectively. Suppose there exist distinct $(x, \alpha), (y, \beta) \in R$
  such that $P^\gamma(x, \alpha) = P^{\gamma}(y, \beta)$. Then from
  Lemma~\ref{lem:sum-equiv} we have for all $(z, \delta) \in C$ that
  \begin{align*}
    S((x, \alpha), (z, \delta)) = L^\gamma_*(P^{\gamma}(x, \alpha), P^{\gamma}(z, \delta)) = L^\gamma_*(P^{\gamma}(y, \beta), P^{\gamma}(z, \delta)) = S((y, \beta), (z, \delta)).
  \end{align*}
  Then $S$ has two identical rows and so $\det(S) = 0$, a contradiction. It
  follows from this and from Lemmas~\ref{lem:M-to-L-homomorphism}
  and~\ref{lem:M-to-L-surjective} that $P^{\gamma}$ embeds $S$ as a submatrix in
  $L^{\gamma}_*$, and so $\rank_p(L^{\gamma}_*) \geq \rank_p(M^{g, \eta}_*)$
\end{proof}

\subsection{Constructing an $\FPR$-Formula}
\label{sec:translating-formulas-to-FPR}
Let $\mathcal{C} := (C_n)_{n \in \nats}$ be a $\PT$-uniform family of
transparent symmetric rank circuits. From Lemma~\ref{lem:transparent-unique} we
may assume without a loss of generality that each $C_n$ is reduced and
injective.

Let $T := \{\AND, \OR, \NOT, \RANK\} \cup \rho \cup \{0,1\}$. It follows from
the Immerman-Vardi theorem and the $\PT$-uniformity of $\mathcal{C}$ that there
is an $\natFP$-interpretation
\begin{align*}
  \Phi := (\phi_G,
  \phi_\Omega, (\phi_{\Sigma, s})_{ s \in T},
  \phi_{\RANK}, (\phi_{\Lambda_R})_{R \in \rho}, \phi_W,
  \phi_{L})
\end{align*}
such that for each $n \in \nats$ when $\Phi$ is interpreted in $\rho$-structure
$\mathcal{A}$ of size $n$ it defines $C_n$ in the number domain. We abuse
notation and also refer to this equivalent circuit as $C_n$. Let $t$ be the
width of this interpretation. Throughout this subsection we use $\mu$, $\nu$,
$\epsilon$, $\eta$, and $\delta$ to denote $t$-length sequences of number
variables and $\kappa$ and $\pi$ to denote individual number variables. We now
describe the formulas in $\Phi$ by describing the relation that each formula
defines when interpreted in a $\rho$-structure $\mathcal{A}$ of size $n$.

\begin{myitemize}
\item $\phi_G(\mu)$ defines the set of gates in the circuit $C_n$ as a subset of
  $[n]^t$. We identify this set with $G_n$, and hence write $G_n \subseteq
  [n]^t$.
\item $\phi_{\Omega}(\kappa_1, \ldots , \kappa_q, \mu)$ is defined such that
  $\mathcal{A} \models \phi_\Omega[a_1, \ldots, a_q, g]$ if, and only if, $g$ is
  a gate, $(a_1, \ldots, a_q) \in [n]^q$, and $\Omega_n(a_1, \ldots, a_q) = g$.
\item $\phi_{\Sigma, s} (\mu)$ is defined for $s \in T$ such that $\mathcal{A}
  \models \phi_s [g]$ if, and only if, $g$ is an input gate and $\Sigma_n (g) =
  s$ or $g$ is an internal gate and $\Sigma_n(g)$ is a function of type $s$.
\item $\phi_{\RANK}(\mu, \eta_1, \eta_2, \delta_1, \delta_2, \delta_3)$ is
  defined such that $\mathcal{A} \models \phi_\RANK[g, p, t, d_1, d_2, d_3]$ if,
  and only if, $\Sigma (g) = \RANK^t_p[d_1, d_2, d_3]$.
\item Let $R \in \rho$. Then $\phi_{\Lambda_R}(\mu, \delta_1, \ldots,
  \delta_{r_R})$ is such that $\mathcal{A} \models \phi_{\Lambda_R} [g, a_1,
  \ldots, a_{r_R}]$ if, and only if, $(a_1, \ldots, a_{r_R}) \in [n]^{r_R}$ and
  $g$ is a relational gate such that $\Sigma (g) = R$ and $(\Lambda_n)_R (g) =
  (a_1, \ldots, a_{r_R})$.
\item $\phi_W(\mu, \nu)$ is defined such that $\mathcal{A} \models \phi_W[g, h]$
  if, and only if, $g$ and $h$ are gates and $h \in H_g$.
\item $\phi_{L}(\mu, \nu, \delta_1, \delta_2, \delta_3 )$ is defined such that
  $\mathcal{A} \models \phi_{L}[g, h, a_1, a_2, a_3]$ if, and only if, $g$ is an
  rank gate such that $R$ is a relation symbol in the vocabulary of
  $\Sigma_n(g)$, $h \in H_g$, and $L_n(g)((a_1, a_2, a_3)) = h$.
\end{myitemize}

The interpretation $\Phi$ does not define a circuit exactly how we might expect.
In particular, it does not contain formulas corresponding to $\Sigma_n$ or
$L_n$. The formula $\phi_L$ only defines $L_n$ for rank gates and $\phi_{\Sigma,
  s}$ does not define $\Sigma_n$ but rather determines to which family of
functions each $\Sigma_n(g)$ belongs. These formulas together suffice to define
the circuit.

By Corollary~\ref{cor:poly-size-support-bound} there are constants $n_0$ and $k$
such that for all $n > n_0$, we have $\SPs{C_n} \leq k$. Notice that for each $n
\leq \max(n_0, 3k)$, there are constantly many bijections from the universe of a
$\rho$-structure $\mathcal{A}$ of size $n$ to $[n]$. It follows that here exists
a $\FPR$-formula that evaluates $C_n$ for any $n \leq \max(n_0, 3k)$ by
explicitly quantifying over all of these constantly many bijections, and then
evaluating the circuit with respect to each bijection. For the rest of this
section we fix an $n > \max(n_0, 3k)$ and let $\mathcal{A}$ denote a
$\rho$-structure of size $n$.

The gates in a circuit and elements of the universe of a gate are encoded as
$t$-length sequences of number variables. In the remainder of this section we
use $\mu$ and $\nu$ to denote gates and $\delta$ and $\epsilon$ to denote
elements of the universe of a gate. We use $\kappa$, $\pi$ and $\lambda$ to
denote single number variables and $\vec{\kappa}$ to denote a $2k$-length tuple
of number variables. We use $\vec{x}$ and $\vec{y}$ to denote $k$-length
sequences of element variables and use $\vec{z}$ to denote $2k$-length sequences
of element variables. We use $U$ and $V$ to denote second-order variables. If
$S$ is a subset of an ordered set we write $\vec{S}$ to denote the $\vert S
\vert$-tuple given by listing the elements of $S$ in order. Let $X$ and $Y$ be
sets, $\vec{a}$ be a sequence in $X$, and $\vec{u}$ be a sequence of distinct
elements in $Y$ such that $\vert \vec{u} \vert \leq \vert \vec{a} \vert$. Let
$\alpha^{\vec{u}}_{\vec{a}}: \img (\vec{u}) \ra X$ be such that
$\alpha^{\vec{u}}_{\vec{a}}(b) := \vec{a} \circ \vec{u}^{-1}(b)$ for all $b \in
\img(u)$.

We aim to give a recursive definition of $\EV_g$ in $\FPR$. We have from
Theorem~\ref{thm:support-theorem-universe} that $\consp(g)$ has size at most
$k$, but it may not be exactly equal to $k$. If $\vert \consp(g) \vert = \ell$,
we define
\begin{align*}
	\overline{\EV}_g := \{ (a_1, \ldots , a_k) \in [n]^k : \alpha^{(a_1 , \ldots , a_\ell)}_{\vec{\consp}(g)} \in \EV_g \text{ and } \forall i, j \in [k] \, \, (i \neq j \implies a_i \neq a_j) \}. 
\end{align*}

More formally then, we aim to define an $\FPR$-formula $\theta(\mu, \vec{x})$
such that $\mathcal{A} \models \theta[g, \vec{a}]$ if, and only if, $\vec{a} \in
\overline{\EV}_g$. We do so by defining for each $s \in T$ a formula $\theta_s$
that gives a recursive definition of $\theta$ for any gate associated with the
symbol $s$. That is, for a second-order variable $V$ with the same type as
$(\mu, \vec{x})$ we define for each $s \in T$ a formula $\theta_s$ so that for
each gate $g$ with $\mathcal{A} \models \phi_{\Sigma, s}[g]$ and each $\vec{a}
\in A^k$, if $V$ is mapped to a relation $\beta(V)$ such that for all $h \in
H_g$, $(h, \vec{b}) \in \beta(V)$, if, and only if, $\vec{b} \in
\overline{\EV}_g$, then $\mathcal{A} \models \theta_s [g, \vec{a}; \beta(V)]$
if, and only if, $\vec{a} \in \overline{\EV}_g$.

Anderson and Dawar~\cite{AndersonD17} have already defined $\theta_s$ for each
$s \in T \setminus \{\RANK\}$. In each of these cases the definition of
$\theta_s$ is very straightforward, and so we reproduce these formulas below
with minimal discussion and minor changes for the sake of presentation.

We first define a few auxiliary formulas. These appear in~\cite{AndersonD17},
and we reproduce these definitions with minor modification. We have from the
Immerman-Vardi theorem and Lemma~\ref{lem:computing-support-orbit} that there
exists a $\natFP$-formula $\FA{supp}(\mu, \kappa)$ such that $\mathcal{A}
\models \FA{supp} [g, u]$ if, and only if, $g$ is a gate and $u \in \consp(g)$.
We can define from $\FA{supp}$ a formula $\FA{supp}_i$ for each $i \in \nats$
such that $\mathcal{A} \models \FA{supp}_i[g, u]$ if, and only if, $u$ is the
$i$th element of $\consp(g)$. We define these formulas by induction as follows
\begin{align*}
  \FA{supp}_1(\mu, \kappa) &:\equiv \FA{supp}(\mu, \kappa) \land (\forall \, \pi (\pi < \kappa) \implies \neg \FA{supp}(\mu, \pi)) \\
  \FA{supp}_{i + 1} (\mu, \kappa) &:\equiv \FA{supp} (\mu, \kappa) \land \exists \pi_1 (\pi_1 < \kappa \land \FA{supp}_i(\mu, \pi_1) \\
                           & \land \forall \pi_2 ((\pi_1 < \pi_2 < \kappa) \implies \neg \FA{supp}(\mu, \pi_2))).
\end{align*}

We can define from these formulas a formula $\FA{agree}(\mu, \nu, \vec{x},
\vec{y})$ such that $\mathcal{A} \models \FA{agree}(g, h, \vec{a}, \vec{b})$ if,
and only if, $g$ and $h$ are gates, $h \in H_g$, and
$\alpha^{\vec{a}}_{\vec{\consp}(g)} \sim \alpha^{\vec{b}}_{\vec{\consp}(h)}$.
This formula is defined as follows
\begin{align*}
  \FA{agree} (\mu, \nu, \vec{x}, \vec{y}) :\equiv \, & \phi_G(\mu) \land \phi_G(\nu) \land \bigwedge_{1 \leq e , d \leq [k]} [\forall \delta \, (\FA{supp}_e(\mu, \delta) \land \FA{supp}_d (\nu, \delta)) \implies x_e = y_d) \land \\
                                                     &\forall \delta_1, \delta_2\, ((\FA{supp}_e (\mu, \delta_1) \land \FA{supp}_d (\mu, \delta_2) \land x_e = x_d) \implies \delta_1 = \delta_2)].
\end{align*}

We define $\theta_s$ for each $s \in T \setminus \{\RANK\}$ as
follows~\cite{AndersonD17}
\begin{align*}
  \theta_0 (\mu, \vec{x}) &:\equiv \exists y \, (y \neq y)\\
  \theta_1 (\mu, \vec{x}) &:\equiv \bigwedge_{1 \leq i < j \leq k} x_i \neq x_j \\
  \theta_R (\mu, \vec{x}) &:\equiv (\bigwedge_{1 \leq i < j \leq k} x_i \neq x_j) \land \exists y_1, \ldots, y_r \exists \kappa_1, \ldots, \kappa_r \, R(y_1, \ldots, y_r) \land  \phi_{\Lambda_R}(\mu, \kappa_1, \ldots, \kappa_{r}) \land \\ & \qquad {}   \bigwedge_{i \in [r]} \bigwedge_{j \in [k]} (\FA{supp}_j (\mu, \kappa_i) \implies y_i = x_j)\\
  \theta_{\OR} (\mu, \vec{x}) &:\equiv (\bigwedge_{1 \leq i < j \leq k} x_i \neq x_j) \land \exists \nu \exists \vec{y}  \, \psi_{W} (\nu, \mu) \land \FA{agree}(\mu, \nu, \vec{x}, \vec{y}) \land V(\nu, \vec{y})\\
  \theta_{\AND} (\mu, \vec{x}) &:\equiv (\bigwedge_{1 \leq i < j \leq k} x_i \neq x_j) \land \forall \nu \forall \vec{y} \, ((\psi_{W} (\nu, \mu) \land \FA{agree}(\mu, \nu, \vec{x}, \vec{y})) \implies V(\nu, \vec{y}))\\
  \theta_{\NOT} (\mu, \vec{x}) &:\equiv (\bigwedge_{1 \leq i < j \leq k} x_i \neq x_j) \land \exists \nu \exists \vec{y}  \, \psi_{W} (\nu, \mu) \land \FA{agree}(\mu, \nu, \vec{x}, \vec{y}) \land \neg V(\nu, \vec{y})
\end{align*}

The remainder of this section is dedicated to defining $\theta_{\RANK}$. We
begin my extending the auxiliary formulas $\FA{supp}$ and $\FA{agree}$ to
supports of elements of the universe of a gate. Let $s \in [3]$. From the
Immerman-Vardi theorem and Lemma~\ref{lem:computing-support-orbit-index} there
is a formula $\FA{supp}^s(\mu, \delta, \kappa)$ such that $\mathcal{A} \models
\FA{supp}^s[g, b, u]$ if, and only if, $g$ is a gate, $b$ is the $s$th element
of the universe of $g$, and $u \in \consp(b)$. We can define for each $j \in
\nats$, using a similar approach as for $\FA{supp}$, a formula
$\FA{supp}^s_j(\mu, \delta, \kappa)$ such that $\mathcal{A} \models
\FA{supp}^s_j[g, u, m]$ if, and only if, $\mathcal{A} \models \FA{supp}^s[g, u,
m]$ and $m$ is the $j$th element of $\consp(u)$. We can use a similar approach
as in the definition $\FA{agree}$ to define for each $s \in [3]$ a formula
$\FA{agree}^s_L (\mu, \nu, \delta, \vec{x}, \vec{y}, \vec{z})$ such that
$\mathcal{A} \models \FA{agree}^s_L [g, h, u, \vec{a}, \vec{b}, \vec{c}]$ if,
and only if, $g$ and $h$ are gates with $h \in H_g$, $u$ is an element of the
$s$th sort of the universe of $g$, and $\alpha^{\vec{a}}_{\vec{\consp}(g)}$,
$\alpha^{\vec{b}}_{\vec{\consp}(h)}$, and $\alpha^{\vec{c}}_{\vec{\consp}(u)}$
are pairwise compatible. Let $\FA{agree}^s_L(\mu, \delta, \vec{x}, \vec{z})
:\equiv \exists \nu \vec{y} \, \FA{agree}^s_L (\mu, \nu, \delta, \vec{x},
\vec{y}, \vec{z})$.

We have from Lemma~\ref{lem:compute-automorphisms-labels} and the Immerman-Vardi
theorem that for each $s \in [3]$ there is a formula $\FA{move}^s(\mu, \delta_1,
\delta_2, \vec{\kappa})$ such that $\mathcal{A} \models \FA{move}^s[g, b_1, b_2,
\vec{u}]$ if, and only if, $g$ is a gate, $b_1$ and $b_2$ are elements of the
$s$th sort of the universe of $g$, for all $a, b \in [2k]$ if $a \neq b$ then
$u_a \neq u_b$, and there exists $\sigma \in \spstab{g}$ such that for all $a
\in [\vert \consp(b_1) \vert]$, $\sigma (\vec{\consp}(b_1)(a)) = u_a$ and
$\sigma (b_1) = b_2$. In other words, $\mathcal{A} \models \FA{move}^s[g, b_1,
b_2, \vec{u}]$ if, and only if, the function that maps the support of $b_1$ to
$\vec{u}$ extends to a permutation in $\spstab{g}$ that maps $b_1$ to $b_2$.

For each $s \in [3]$ let $\FA{orb}^s(\mu, \delta_1, \delta_2) :\equiv \exists
\vec{\kappa} \, \FA{move}^s(\mu, \delta_1, \delta_2, \vec{\kappa})$. It can be
seen that $\mathcal{A} \models \FA{orb}^s [g, b_1, b_2]$ if, and only if, $b_1$
and $b_2$ are elements of the $s$th sort of the universe of $g$, and $b_1 \in
\orb(b_2)$. Let
\begin{align*}
  \FA{min-orb}^s (\mu, \delta) :\equiv \forall \epsilon \, (\FA{orb}^s (\mu, \delta, \epsilon) \implies \delta \leq \epsilon). 
\end{align*}

We next define a series of $\FPR$-formulas that implement the definition of
$M^{g, \alpha}$ given in Section~\ref{sec:recursive-matrix} for a rank gate $g$
and assignment $\alpha \in A^{\underline{\consp}(g)}$. We first need to define
$\bar{J}^{g, \alpha}$ and $J^{g, \alpha}$ in the logic. The definition of
$\bar{J}^{g, \alpha}$ given in Section~\ref{sec:recursive-matrix} can be
directly implemented in an obvious manner to define an $\natFP$-formula
$\psi_{\bar{J}} (\mu, \vec{x}, \delta_{1}, \vec{z}_1, \delta_{2}, \vec{z}_2,
\delta_{3} \vec{z}_3, \vec{\kappa}_1, \vec{\kappa}_2, \vec{\kappa}_3)$ such that
$\mathcal{A} \models \psi_{\bar{J}}[g, \vec{x}, u_1, \vec{c}_1, u_2, \vec{c}_2,
u_3, \vec{c}_3, \vec{m}_1, \vec{m}_2, \vec{m}_3]$ if, and only if, for $\alpha
:= \alpha^{\vec{a}}_{\vec{\consp}(g)}$,
\begin{align*}\bar{J}^{g,
  \alpha} ((u_1,
  \alpha^{\vec{c}_1}_{\vec{\consp}(u_1)}), (u_2,
  \alpha^{\vec{c}_2}_{\vec{\consp}(u_2)}), (u_3,
  \alpha^{\vec{c}_3}_{\vec{\consp}(u_3)})) = (\bar{\sigma}_1, \bar{\sigma}_2,
  \bar{\sigma}_3),\end{align*} where for each $s \in [3]$, $\bar{\sigma}$ maps
$\vec{\consp}(u_s)$ to an initial segment of $\vec{\kappa}_s$.

We now use $\psi_{\bar{J}}$ to define an $\natFP$-formula $\psi_{J} (\mu,
\vec{x}, \delta_{1}, \vec{z}_1, \delta_{2}, \vec{z}_2, \delta_{3}, \vec{z}_3,
\nu, \vec{y})$ such that $\mathcal{A} \models \psi_{J}[g, \vec{a}, u_1,
\vec{c}_1, u_2, \vec{c}_2, u_3, \vec{c}_3, h, \vec{b} \,]$ if, and only if, $g$
is an internal gate, $h \in H_g$, and for $\alpha :=
\alpha^{\vec{a}}_{\vec{\consp}(g)}$ we have $J^{g, \alpha} ((u_1,
\alpha^{\vec{c}_1}_{\vec{\consp}(u_1)}), (u_2,
\alpha^{\vec{c}_2}_{\vec{\consp}(u_2)}), (u_3,
\alpha^{\vec{c}_3}_{\vec{\consp}(u_3)})) = (h,
\alpha^{\vec{b}}_{\vec{\consp}(h)})$. This formula is defined as follows
\begin{align*}
  \psi_{J} (\mu,
  \vec{x}, \delta_{1}, \vec{z}_1, \delta_{2}, \vec{z}_2, \delta_{3}, \vec{z}_3,
  \nu, \vec{y}) :\equiv \,  & \exists \vec{\kappa}_1, \vec{\kappa}_2, \vec{\kappa}_3 [\psi_{\bar{J},} (\mu, \vec{x}, \delta_1, \vec{z}_{1}, \delta_2, \vec{z}_{2}, \delta_3, \vec{z}_{3}, \vec{\kappa}_1, \vec{\kappa}_2, \vec{\kappa}_3) \land \\ &\exists \delta_1', \delta_2', \delta_3'(\phi_{L}(\mu, \nu, \delta_1', \delta_2', \delta_3') \land \\ & \exists \vec{z}_1', \vec{z}_2', \vec{z}_3' [\bigwedge_{j \in [3]} [\FA{move}^{j} (g, \delta_j, \delta_j', \vec{\kappa}_j) \land \FA{agree}^{j}_L(\mu, \nu, \delta_j', \vec{x}, \vec{y}, \vec{z}_j')\land\\ &  [\bigwedge_{1 \leq a < b \leq 2k} \vec{z}_j'(a) \neq \vec{z}_j'(b) \land \bigwedge_{a \in [2k]} (\forall \lambda \, (\neg \FA{supp}^{j}(\mu, \delta', \lambda))) \lor \\ & \bigvee_{b \in [2k]} \FA{supp}^{j}_{a} (\mu, \delta', \vec{\kappa}_j(b)) \land \vec{z}_j'(a) = \vec{z}_j (b)))]]]
\end{align*}

We now define for $s \in [3]$ a formula $\psi^D_s$ that defines $I^{g,
  \alpha}_s$, the $s$th sort in the domain of $M^{g, \alpha}$. For $s \in [3]$
let
\begin{align*}
  \psi^{D}_s (\mu, \vec{x}, \delta, \vec{z}) :\equiv \FA{min-orbit}^s (\mu, \delta) \land \FA{agree}^s_L (\mu, \delta, \vec{x}, \vec{z}).
\end{align*}

We now define a formula $\psi_M$ that inductively defines for a rank gate $g$
and assignment $\alpha \in A^{\underline{\consp}(g)}$ a function $M$ which is
equal to $M^{g, \alpha}$ when restricted to the domain of $M^{g, \alpha}$ and is
$0$ everywhere else. These additional $0$s have no effect on the rank. More
precisely, the formula $\psi_{M} (\mu, \vec{x}, \delta_1, \vec{z}_1, \delta_2,
\vec{z}_2, \delta_3, \vec{z}_3; V)$ is defined such that $\mathcal{A} \models
\psi_{M} [g, \vec{a}, u_1, \vec{c}_1, u_2, \vec{c}_2, u_3, \vec{c}_3; \beta(V)]$
if, and only if, for all $s \in [3]$, $(u_s,
\alpha^{\vec{c}_s}_{\vec{\consp}(u_s)}) \in I^{g, \alpha}_s$, where $\alpha :=
\alpha^{\vec{a}}_{\vec{\consp}(g)}$, and if
\begin{align*}
  J^{g, \alpha}((u_1, \alpha^{\vec{c}_1}_{\vec{\consp}(u_1)}), (u_2,
  \alpha^{\vec{c}_2}_{\vec{\consp}(u_2)}), (u_3,
  \alpha^{\vec{c}_3}_{\vec{\consp}(u_3)})) = (h, \eta)
\end{align*} and $\beta(V)$ is an
assignment to $V$ such that for all $h \in H_g$ and $\vec{b} \in
A^{\underline{k}}$, $(h, \vec{b}) \in \beta(V)$ if, and only if, $\vec{b} \in
\overline{\EV}_h$ then $\eta \in \EV_h$. This formula is defined as follows
\begin{align*}
  \psi_{M}(\mu, \vec{x}, \delta_1, \vec{z}_1,  \delta_2, \vec{z}_2, \delta_3, \vec{z}_3; V) :\equiv &\bigwedge_{s \in [3]} \psi^D_s (\mu, \vec{x}, \delta_s, \vec{z}_s) \land \\ & \exists \nu , \vec{y} \, [\psi_{J}(\mu, \vec{x}, \delta_1, \vec{z}_1,  \delta_2, \vec{z}_2, \delta_3, \vec{z}_3, \nu, \vec{y}) \land V(\nu, \vec{y})].
\end{align*}

We now aim to define a formula $\psi^M_*$ that defines a structure similar to
$M^{g, \alpha}_*$ but with some additional all-zero rows and columns. This
structure gives a matrix that has the same rank as $M^{g, \alpha}_*$ and so
suffices for evaluating $g$. We first define a number term
$\zeta_{\text{oc}}(\mu, \delta)$ such that if $g$ is a rank gate and $u$ is an
element of the third sort of $g$ then $\zeta_{\text{oc}}[g, u]$ evaluates to
$\vert \orb_{\stab(u)}(\consp(u))\vert$. This formula is defined as follows

\begin{align*}
  \zeta_{\text{oc}}(\mu, \delta) :\equiv \#_{\vec{\kappa}} [\FA{move}^3(\mu, \delta, \delta, \vec{\kappa}) \land \bigwedge_{i \in [2k]}\FA{supp}^3(\mu, \delta, \kappa_i)].
\end{align*}

The number term $\zeta_{\text{oc}}$ and the formula $\psi_{M}$ suffice to define
the numerators and denominators in Equation~\ref{eq:M-star}. The divisions and
additions can be trivially expressed in $\natFP$, and so we can define a number
term $\zeta^M_* (\mu, \vec{x}, \delta_1, \vec{z}_1, \delta_2, \vec{z}_2; V)$
such that if $g$ is a rank gate, $\vec{a} \in A^{\underline{k}}$, $u_1$ is an
element of the first sort of $g$, $u_2$ is an element of the second sort of $g$,
and $\vec{b}_1, \vec{b}_2 \in A^{\underline{2k}}$ are such that
$\alpha^{\vec{b}_1}_{\vec{\consp}(u_1)}$ and
$\alpha^{\vec{b}_2}_{\vec{\consp}(u_2)}$ are compatible with
$\alpha^{\vec{a}}_{\vec{\consp}(g)}$, then $\zeta^M_* [g, \vec{a}, u_1,
\vec{b}_1, u_2, \vec{b}_2]$ evaluates to $M^{g,
  \alpha^{\vec{a}}_{\vec{\consp}(g)}}_* ((u_1,
\alpha^{\vec{b_1}}_{\vec{\consp}(u_2)}), (u_1,
\alpha^{\vec{b_2}}_{\vec{\consp}(u_2)})$, and otherwise $\zeta^M_* [g, \vec{a},
u_1, \vec{b}_1, u_2, \vec{b}_2]$ evaluates to $0$.

The number term $\zeta^M_*$ for a gate $g$ and assignment $\alpha \in
A^{\underline{\consp(g)}}$ (written as a $k$-tuple) defines a matrix $M$ such
that $\rank_p (M) = \rank_p(M^{g, \alpha})$ for any prime $p$. It follows from
Theorem~\ref{thm:rank-equivilence} that $\rank_p (M) = \rank (L^\gamma_*)$.

We define $\theta_{\RANK} (\mu, \vec{x}; V)$ as follows
\begin{align*}
  \theta_{\RANK} (\mu, \vec{x}; V) :\equiv &(\bigwedge_{1 \leq i < j
                                             \leq k} x_i \neq x_j) \land \exists \epsilon_1, \epsilon_2, \eta_1, \eta_2, \eta_3 [\phi_{\RANK}(\mu, \epsilon_1, \epsilon_2, \eta_1, \eta_2, \eta_3) \land \\ &[\rank (\vec{z}_1\delta_1 \leq \eta_1, \vec{z}_2\delta_2 \leq \eta_2, \epsilon_2) . \zeta^M_* (\mu, \vec{x}, \delta_1, \vec{z}_1, \delta_2, \vec{z}_2)]
\end{align*}

We now define $\theta(\mu, \vec{x})$ as follows
\begin{align*}
  \theta (\mu, \vec{x}) :\equiv [\ifp_{V,\nu \vec{y}} \bigvee_{s \in T} (\phi_{\Sigma, s}(\mu) \land \theta_s (\nu, \vec{y} ))] (\mu, \vec{x}).
\end{align*}

We finally define the $\FPR$-formula $Q$ that defines the same $q$-ary query as
$\mathcal{C}$. The definition of this formula is similar to one given
in~\cite{AndersonD17}. Let
\begin{align*}
  Q (y_1, \ldots y_q)  :\equiv & \, \exists \vec{x} \, \exists \mu, \kappa_1 , \ldots,  \kappa_q \pi_1 , \ldots , \pi_k \, [\theta (\mu, \vec{x}) \land \phi_\Omega (\kappa_1, \ldots, \kappa_q, \mu) \land \\
                               & \bigwedge_{1 \leq i \leq k} (\FA{supp}_i (\mu, \pi_i) \lor \forall \pi \, (\neg \FA{supp}_i (\mu, \pi))) \land \\
                               & \bigwedge_{1 \leq i \leq k} \bigwedge_{1 \leq j \leq q}((\FA{supp}_i (\mu, \pi_i) \land (x_i = y_j)) \implies \kappa_j = \pi_i) \land \\ &                                                                                                                                                        \bigwedge_{1 \leq j \leq q} \bigvee_{1 \leq i \leq k} (x_i = y_j \land \FA{supp}_i (\mu, \pi_i))].
\end{align*}
We could informally understand the formula $Q$ as inverting the assignment
denoted by $(y_1, \ldots, y_q)$, selecting the corresponding output gate, and
then evaluating this output gate. This completes the proof of
Theorem~\ref{thm:circuits-formulas} which, combined with
Theorem~\ref{thm:translating-formulas-to-circuits}, completes the proof
Theorem~\ref{thm:main-result}.

\end{document}

\section{Concluding Discussion}

$\FPR$ is one of the most expressive logics we know that is still contained in
$\PT$ and understanding its expressive power is an important question. The main
result of this paper establishes an equivalence between the expressive power of
$\FPR$ and the computational power of uniform families of transparent symmetric
rank circuits. Not only does this establish an interesting characterization of
an important logic, it also deepens our understanding of the connection between
logic and circuit complexity and sheds new light on foundational aspects of the
circuit model.

The circuit characterization helps emphasise certain important aspects of the
logic. Given that $\PT$-uniform families of invariant circuits (without the
restriction to symmetry) express all properties in $\PT$, we can understand the
inability of $\FPC$ (and, conjecturally, $\FPR$) to express all such properties
as essentially down to symmetry. As with other (machine) models of computation,
the translation to circuits exposes the inherent combinatorial structure of an
algorithm. In the case of logics, we find that a key property of this structure
is its symmetry and the translation to circuits provides us with the tools to
study it.

Still, the most significant contribution of this paper is not in the main result
but in the techniques that are developed to establish it, and we highlight some
of these now. The conclusion of~\cite{AndersonD17} says that the support theorem
is ``largely agnostic to the particular [\ldots] basis'', suggesting that it
could be easily adapted to include other gates. This turns out to have been a
misjudgment. Attempting to prove the support theorem for a basis that includes
rank threshold gates showed us the extent to which both the proof of the theorem
and, more broadly, the definitions of circuit classes, rest heavily on the
assumption that all functions computed by gates are symmetric. Thus, in order to
define what the ``symmetry'' condition might mean for circuits that include rank
threshold gates, we radically generalise the circuit framework to allow for
gates that take structured inputs (rather than sets of $0$s and $1$s) and are
invariant under isomorphisms. This leads to a refined notion of circuit
automorphism, which allows us to formulate a notion of symmetry and prove a
version of the support theorem. Again, in that proof, substantial new methods
are required.


The condition of \emph{transparency} makes the translation of uniform circuit
families into formulas of logic (which is the difficult direction of our
characterisation) possible, but it complicates the other direction. Indeed, the
natural translation of formulas of $\FPR$ into uniform circuit families yields
circuits which are symmetric, but not transparent. This problem is addressed by
introducing gadgets in the translation---which for ease of exposition, we did in
formulas of $\FOrk$ which are then translated into circuits in the natural way.
Thus, the restriction to transparent circuits is sufficient to get both
directions of the characterisation.

In short, we can represent the proof of our characterisation through the three
equivalences in this triangle.

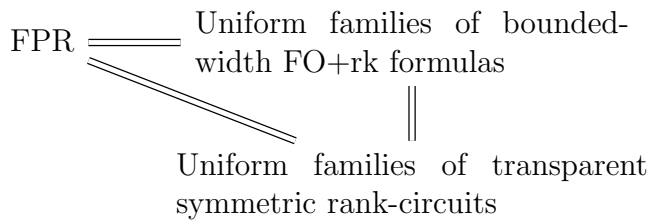
\begin{center}
  \begin{tikzcd}
    \FPR \ar[r, equal] \ar[dr, equal] & \parbox{0.35\textwidth}{Uniform families
      of
      bounded-width $\FOrk$ formulas}  \ar[d, equal]\\
    & \parbox{0.38\textwidth}{Uniform families of transparent symmetric
      rank-circuits}
  \end{tikzcd}
\end{center}

This highlights another interesting aspect of our result. The first translation,
of $\FPR$ to uniform families of $\FOrk$ formulas was given
in~\cite{Dawar09logicswith} and used there to establish arity lower bounds.
However, this was for a weaker version of the rank logic rather than the
strictly more expressive one defined by Gr\"{a}del and Pakusa~\cite{GradelP15a}.
The fact that we can complete the cycle of equivalences with the more powerful
logic demonstrates that the definition of Gr\"{a}del and Pakusa is the ``right''
formulation of $\FPR$.

\subsection*{Future Work}
There are many directions of work suggested by the methods and results developed
in this paper. First of all, there is the question of transparency. We introduce
it as a technical device that enables our characterisation to go through. Could
it be dispensed with? Or are $\PT$-uniform families of transparent symmetric
rank-circuits strictly weaker than families without the restriction of
transparency?

The framework we have developed for working with circuits with structured inputs
is very general and mostly not specific to rank gates. In his thesis
Wilsenach~\cite{wilsenachthesis2019} introduces \emph{generalised operators},
which generalise Lindstr\"om quantifiers as well as counting and rank operators,
and establishes a general correspondence analogous to
Theorem~\ref{thm:main-result} between fixed-point logics extended with
generalised operators and symmetric circuits over an appropriate basis. It
remains an open question how far this correspondence can be generalised. For
example, can we use this framework to develop a circuit characterisation of
$\CPTC$?






At the moment, we have little by way of methods for proving inexpressibility
results for $\FPR$, whether we look at it as a logic or in the circuit model.
The logical formulation lays emphasis on some parameters (the number of
variables, the arity of the operators, etc.) which we can treat as resources
against which to prove lower bounds. On the other hand, the circuit model brings
to the fore other, more combinatorial, parameters. One such is the fan-in of
gates and a promising and novel approach is to try and prove lower bounds for
symmetric circuits with gates with bounded fan-in. We might ask if it is
possible to compute $\AND[3]$ using a symmetric circuit with gates that have
fan-in two. Perhaps we could also combine the circuit view with lower-bound
methods from logic, such as pebble games. Dawar~\cite{Dawar2016} has shown how
the bijection games of Hella~\cite{Hella19961} can be used directly to prove
lower bounds for symmetric circuits without reference to the logic. We also have
pebble games for $\FPR$~\cite{DawarH2012}, and it would be interesting to know
if we can use these on circuits and how the combinatorial parameters of the
circuit interact with the game.

Finally, we note that some of the interesting directions on the interplay
between logic and symmetric circuits raised in~\cite{AndersonD17} remain
relevant. Can we relax the symmetry condition to something in between requiring
invariance of the circuit under the full symmetric group (the case of symmetric
circuits) and requiring no invariance condition at all? Can such restricted
symmetries give rise to interesting logics in between $\FPR$ and $\PT$?

\section*{Acknowledgements}
The work reported here has benefitted enormously from discussions we have had
with a number of fellow researchers. We would particularly like to thank Joanna
Fawcett, who pointed us to Theorem~\ref{thm:dixonmort} and Matt Anderson and
Martin Otto with whom we had extensive discussions on symmetric circuits when we
were all visitors at the Simons Institute for the Theory of Computing at
Berkeley in 2016. In particular, Martin provided us with a note and an
explanation of how his methods could be used to prove a version of the support
theorem (see Remark~\ref{rem:support-theorem-otto} for details). He also served
as an examiner for the second author's dissertation and his comments and
suggestions given then have greatly improved this paper. For all this, we are
extremely grateful.

\bibliographystyle{plain} \bibliography{references.bib}

\begin{thebibliography}{10}

\bibitem{AndersonD17}
M.~Anderson and A.~Dawar.
\newblock On symmetric circuits and fixed-point logics.
\newblock {\em Theory of Computing Systems}, 60(3):521--551, 2017.

\bibitem{BLASS1999141}
Andreas Blass, Yuri Gurevich, and Saharon Shelah.
\newblock Choiceless polynomial time.
\newblock {\em Annals of Pure and Applied Logic}, 100(1):141 -- 187, 1999.

\bibitem{Dawar-siglog}
A.~Dawar.
\newblock The nature and power of fixed-point logic with counting.
\newblock {\em ACM SIGLOG News}, 2(1):8--21, 2015.

\bibitem{Dawar2016}
A.~Dawar.
\newblock On symmetric and choiceless computation.
\newblock In Mohammad~Taghi Hajiaghayi and Mohammad~Reza Mousavi, editors, {\em
  Topics in Theoretical Computer Science}, pages 23--29, Cham, 2016. Springer
  International Publishing.

\bibitem{Dawar09logicswith}
A.~Dawar, M.~Grohe, B.~Holm, and B.~Laubner.
\newblock Logics with rank operators.
\newblock In {\em 2009 24th Annual IEEE Symposium on Logic In Computer Science
  (LICS)}, pages 113--122, 2009.

\bibitem{DawarH2012}
A.~Dawar and B.~Holm.
\newblock Pebble games with algebraic rules.
\newblock In Artur Czumaj, Kurt Mehlhorn, Andrew Pitts, and Roger Wattenhofer,
  editors, {\em Automata, Languages, and Programming}, pages 251--262, Berlin,
  Heidelberg, 2012. Springer Berlin Heidelberg.

\bibitem{DawarW18}
A.~Dawar and G.~Wilsenach.
\newblock Symmetric circuits for rank logic.
\newblock In {\em 27th {EACSL} Annual Conference on Computer Science Logic,
  {CSL} 2018, September 4-7, 2018, Birmingham, {UK}}, pages 20:1--20:16, 2018.

\bibitem{DawarW20}
Anuj Dawar and Gregory Wilsenach.
\newblock Symmetric arithmetic circuits.
\newblock In {\em 47th International Colloquium on Automata, Languages, and
  Programming, {ICALP} 2020}, pages 36:1--36:18, 2020.

\bibitem{DENENBERG1986216}
L.~Denenberg, Y.~Gurevich, and S.~Shelah.
\newblock Definability by constant-depth polynomial-size circuits.
\newblock {\em Information and Control}, 70(2):216--240, 1986.

\bibitem{dixon1996permutation}
J.D. Dixon and B.~Mortimer.
\newblock {\em Permutation Groups}.
\newblock Graduate Texts in Mathematics. Springer New York, 1996.

\bibitem{GradelP15a}
E.~Gr{\"{a}}del and W.~Pakusa.
\newblock Rank logic is dead, long live rank logic!
\newblock In {\em 2015 24th Annual Conference on Computer Science Logic,
  (CSL)}, pages 390--404, 2015.

\bibitem{grohe2017descriptive}
M.~Grohe.
\newblock {\em Descriptive Complexity, Canonisation, and Definable Graph
  Structure Theory}.
\newblock Lecture Notes in Logic. Cambridge University Press, 2017.

\bibitem{Hella19961}
L.~Hella.
\newblock Logical hierarchies in ptime.
\newblock {\em Information and Computation}, 129(1):1 -- 19, 1996.

\bibitem{Holm2010}
B.~Holm.
\newblock {\em Descriptive complexity of linear algebra}.
\newblock University of Cambridge, 2010.

\bibitem{immerman1999descriptive}
N.~Immerman.
\newblock {\em Descriptive Complexity}.
\newblock Graduate texts in computer science. Springer New York, 1999.

\bibitem{Kolaitis1992}
P.~Kolaitis and M.~Vardi.
\newblock Infinitary logics and 0–1 laws.
\newblock {\em Information and Computation}, 98(2):258 -- 294, 1992.

\bibitem{Otto1997}
M.~Otto.
\newblock The logic of explicitly presentation-invariant circuits.
\newblock In {\em 1996 10th International Workshop, Annual Conference on
  Computer Science Logic (CSL)}, pages 369--384. Springer, Berlin, Heidelberg,
  1997.

\bibitem{wilsenachthesis2019}
Gregory Wilsenach.
\newblock {\em Symmetric {Circuits} and {Model}-{Theoretic} {Logics}}.
\newblock Thesis, University of Cambridge, October 2019.
\newblock Accepted: 2019-10-14T08:50:08Z.

\end{thebibliography}
\end{document}